\documentclass[aps,prl,twocolumn,superscriptaddress]{revtex4-1}
\usepackage{graphicx}  % needed for figures
\usepackage{amsmath} % needed for splitting equations into lines

\usepackage{graphicx}
\usepackage{amsmath,amssymb,amsfonts}
\usepackage{textcomp}
\usepackage{hyperref}
\usepackage{gensymb}
\usepackage{verbatim}

\hypersetup{breaklinks=true,colorlinks=true,urlcolor=black}
\usepackage{color}
\usepackage{soul}
\usepackage{gensymb}

\DeclareGraphicsExtensions{.jpg,.pdf,.png,.eps}

\begin{document}

%\preprint{}

\title{Quantum Formulation of Chiral Vortical Effect in Weyl Semi-metals}



\author{B. Q. Song}
%\affiliation{Ames Laboratory, Iowa State University, Ames, Iowa 50011, USA}
%\affiliation{Department of Physics and Astronomy, Iowa State University, Ames, Iowa 50011, USA}
\affiliation{Department of Physics, University of Houston, Houston, Texas 77204, USA}
\affiliation{Texas Center for Superconductivity, University of Houston, Houston, Texas 77204, USA}
%\author{J. D. H. Smith}
%\affiliation{Ames Laboratory, Iowa State University, Ames, Iowa 50011, USA}
%\affiliation{Department of Mathematics, Iowa State University, Ames, Iowa 50011, USA}
%\author{Y. X. Yao}
%\affiliation{Ames Laboratory, Iowa State University, Ames, Iowa 50011, USA}
%\affiliation{Department of Physics and Astronomy, Iowa State University, Ames, Iowa 50011, USA}
%\author{J. Wang}
%\affiliation{Ames Laboratory, Iowa State University, Ames, Iowa 50011, USA}
%\affiliation{Department of Physics and Astronomy, Iowa State University, Ames, Iowa 50011, USA}


\date{\today}

\begin{abstract}
The chiral vortical effect (CVE) is the generation of an axial current in a rotating Weyl fermion; its description is presently based on semiclassical frameworks. In this work, we develop a fully quantum formulation for CVE, solving the exact evolution of microscopic spinful wavefunctions, which enables a bottom-up quantitative test of semi-classical theories and postulated distributions $f_{\text{CVE}}$ in different reference frames. Notably, it shows that $f_{\text{CVE}}$ is over a ground-state-free Floquet spectrum, qualitatively distinct from a thermal equilibrium distribution (i.e., fermi form $f_F$), underscoring CVE as a non-equilibrium phenomenon, distinguished from other chiral transports. The $f_F$ only approximately holds when three conditions are simultaneously fulfilled: (1) slow rotation $\omega R/v_F\ll 1$, (2) high chemical potential $\mu/(\hbar v_F R)\gg 1$, (3) isotropic symmetry, where $R$ is the size, $v_F$ is fermi velocity. In these conditions, the theory recovers established semiclassical results, including the current-response coefficients and the magnetization contribution; otherwise, it uncovers quantum phenomena such as ``void states", deviation from the semiclassical formula $j_{\text{CVE}} \sim \mu^2$, a $v_F$-independent charge pumping. The theory is based on semimetals, providing more experimentally accessible detection than fundamental Weyl particles.
% at zero temperature.
\end{abstract}

%\keywords{}


\maketitle

\section{I. Introduction}
The chiral vortical effect (CVE) refers to that a rotating Weyl Fermion will spontaneously shift its location, forming an axial current \cite{1,2,3,4,5,6,7,8,8a} -- A salient feature absent in non-chiral fermions \cite{9,10,11} that has motivated studies on broader vorticity-induced chiral effects \cite{12,13,14,15,16}. While the CVE emerges from physical settings \cite{1,3,4} such as neutrino-antineutrino systems \cite{17}, where quantum effects might be important, its description is currently confined to semiclassical frameworks \cite{1,2,3,4,5,8,18,19,20,20a,21,22,23,24}, retaining classical concepts, including force and phase space $\lbrace \boldsymbol{p}, \boldsymbol{x} \rbrace$, in description of states and dynamics. Despite the merits such as in revealing the role of geometry in CVE \cite{14,22,24,25,26,26a}, semi-classical approaches (e.g., wave-packets \cite{27,28,29,30}, kinetic theory \cite{18,22,23,31,32}, hydrodynamics \cite{7,19,33,34,35,36}) lack a unified framework that allows coherent and unbiased comparison among protocols. Moreover, the semi-classical picture might conceal the microscopic roots of CVE distinct from other chiral transports \cite{26,37,38,39,40,41,42,43,44,45,46,47,48}, as well as the unique roles of non-equilibrium, the nature of magnetic contribution ${\nabla}{\times}\boldsymbol{M}$ \cite{20,22}, etc.

However, revising the classical theory into a quantum one is more than defining a Hamiltonian \cite{1,2}, but a series of changes on procedures and notions related to dynamics, extracting observables etc., posing great challenges for a full quantum account for CVE.

First, it involves a transition from a (real) distribution $f(\boldsymbol{p},\boldsymbol{x})$ in phase space $\lbrace\boldsymbol{p},\boldsymbol{x}\rbrace$ \cite{49,50,51,52} to a wavefunction $\varphi$ in Hilbert space -- unlike $f$, the $\varphi$ is generally complex (phase-resolved), may carry spin degrees of freedom, must respect uncertainty principle, and can be subject to quantization.

Second, the equations of dynamics need to be updated. Solving $\varphi(t)$ (or the density operator $\hat{f}(t)$) follows the standard rules of quantum evolution, which are distinct from kinetic equations \cite{18,31,32} or force-based procedures \cite{8,22}. 

Third, the quantized energy for a rotating Weyl fermion proves unbounded (see below) -- the open system lacks a ground state. This poses questions about whether equilibrium-like $f$ is attainable \cite{1,2,13,18,19,20,20a,21,22,23,24,26}. The unbounded spectrum arises from Floquet properties \cite{52,53,54,55,56} and underscores the role of (meta)stable configuration in forming the CVE. 

Fourth, there is a shift in the notion of observables. In addition to observables becoming operators \cite{57,58,59,60}, the ``causality" also changes. In a semi-classical framework, CVE current $\boldsymbol{j}$ is a ``consequence" induced by ``causes", such as fields of velocity $\boldsymbol{v}$ \cite{18,20,23}, magnetization $\boldsymbol{M}$ \cite{19,20a,21,22,24}. In quantum, the $\boldsymbol{j}$, $\boldsymbol{M}$, $\boldsymbol{v}$ all become observables of equal status, which are commonly driven by wavefunction $\varphi$ (or density operator $\hat{f}$) \cite{57,58}. Thus, a ``deriving" relationship becomes correlation between observables subject to a quantum state. 

Fifth, the mesoscopic complexity. The CVE involves non-uniformity in larger scales \cite{60a}, such as the magnetic contribution $\nabla_{x}{\times}\boldsymbol{M}$ \cite{20,22,53a}, which involves ``coarse-grained" spatial derivative (similar to the $\nabla_x$ in Maxwell equations), distinct from the microscopic $\nabla$ in the quantum momentum operator \cite{57}. Thus, the model needs to handle coarse scales. 

These challenges have hindered semiclassical theories from progressing further toward a fully quantum description and should therefore be addressed in building the theory. For experimental relevance, we consider a Weyl semimetal model \cite{10,11,37,38} (instead of an isolated Weyl fermion \cite{1,2,17}), where the chiral fermion corresponds to an imbalance between two opposite Weyl nodes. The model can readily incorporate particle–hole carriers and interband transitions \cite{26a,40,61,62}.

This work pursues three main objectives. First, it establishes a quantum framework for understanding the CVE. Second, it recovers the established semiclassical results, including the induced axial transport \cite{23,53a} and non-uniform magnetization \cite{20,22}. Third, it identifies the regime of validity of the semiclassical theory and uncovers phenomena that lie beyond its scope.

Section 2.1 introduces the physical principles that the quantum Hamiltonian $H$ should obey, presents its specific form, and derives the spectrum and wavefunctions. Section 2.2 is devoted to the invariance of the formulation under reference frame transformation. Section 2.3 discusses the extraction of observables, with particular explanation on the modified causal structure that emerges in quantum frameworks. In Secs. 3.1, 3.2, we apply the quantum theory to reproduce the established response coefficients for charge transport and magnetization, including the treatment of mesoscopic observables. Finally, Secs. 4.1-4.7 are devoted to insights and phenomena that extend beyond the semiclassical description. Appendix B gives a list of notations. 

\section{II. Quantum formulation for CVE}
Relativistic quantum Hamiltonians \cite{1,2} were proposed for CVE in neutrinos, but they do not yet form a complete quantum formulation. First, $\hat{f}(t)$ was assumed to be in equilibrium rather than derived from the (relativistic) quantum equations of motion. Second, the invariance of thermal equilibrium $\hat{f}$ under Lorentz transformations has not been established. 

Here, we consider a non-relativistic quantum $H$ defined on crystal bands, following standard quantum procedures for evolution and observable. It respects invariance under changes of reference frame, which in present case reduces to the more tractable Galilean transformations.

The quantum setup will be introduced from three aspects: (1) the Hamiltonian and its quantized spectrum (distinct from $H$ in \cite{1,2}), (2) the quantum protocol and reference frame transformation under it, (3) extraction of observables. 

\subsection{1. Quantum Hamiltonian}
\begin{figure}
\centering
\includegraphics[scale=0.41]{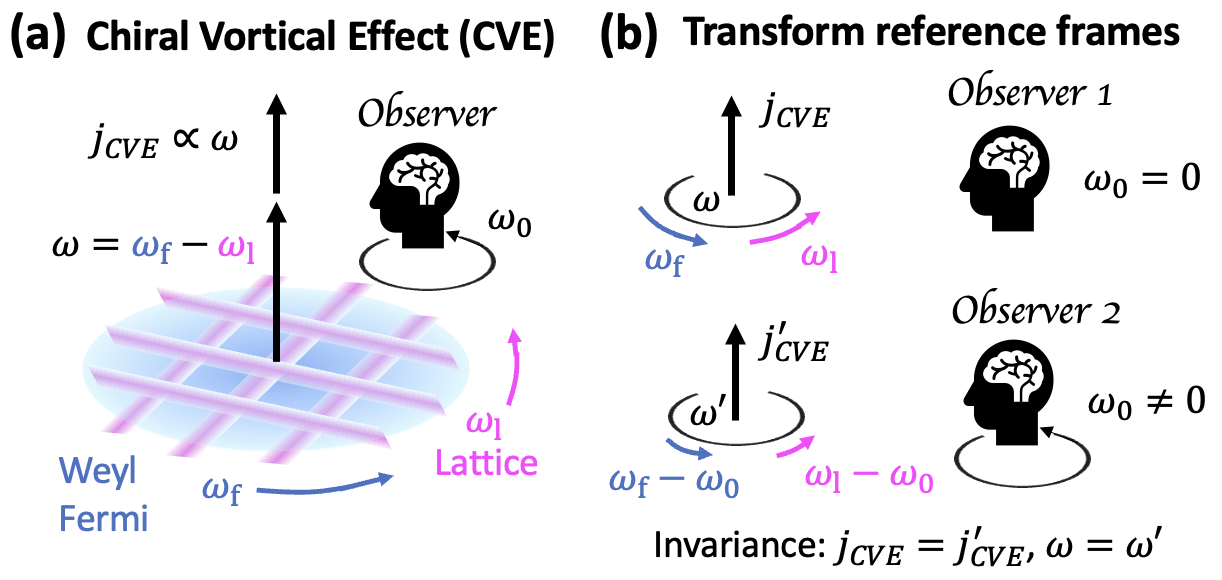}
\caption{\label{fig:epsart2}(color online): (a) A current $j_{\text{CVE}}$ induced by relative rotations between lattice $\omega_{\text{l}}$ and chiral fermion $\omega_{\text{f}}$. (b) The $j_{z}$ and $\omega$ are invariant with $\omega_0$, while $\omega_{\text{l}}$ and $\omega_{\text{f}}$ are variant. \label{f1}}
\end{figure}

The CVE in Weyl semi-metal refers to the axial transport when fermion liquid rotates with respect to the lattice. If the lattice and fermion rotate together (equivalent to rotating the observer but no action acting on the system, Fig.~\ref{f1}(a)), there is no effect. Thus, two angular speeds should be distinguished: one is the \textit{relative} rotation $\omega$ between the lattice and fermion liquid; the other is the observer's rotation $\omega_0$ (with respect to the lab). Since $\omega$ is the difference of two angular velocities, it is invariant with $\omega_0$.

\textbf{Principles for quantum Hamiltonian}. A valid Hamiltonian needs to respect several model-independent physical principles.

1. Observables such as current $j_z$, energy spectrum of $H$, should be invariant with reference frame (rotational) transformation $\omega_0$. In mechanics (whether Newtonian or relativistic), it holds the principle of the $z$-axis speed being independent of the $x$-$y$ motion of the observer \cite{63}. %Thus, for both classical and quantum, as a general requirement, a CVE formulation should be invariant under (axial) reference frame transformations. 
In quantum, the axial symmetry is formulated by $[H,J_z]=0$, serving as a necessary condition for $H$.

2. The axial current $j_z$ computed from $H$ is \textit{not} invariant with $\omega$. In fact, $j_z$ should be dominated by linear $\omega$ -- the basic connotation of CVE \cite{1,2,3,4,5}.

3. The Hamiltonian $H$ is established on a Hilbert space of the Fermi sub-system. The lattice or other environments are treated as potentials. (Otherwise, for instance, given fermion in a optical lattice, the full wavefunction needs to be a combination of bosons and fermions)

4. The Floquet symmetry $H(t)=H(t+2\pi/{\omega}_{\text{l}})$ \cite{53,54,55}, where $\omega_{\text{l}}$ is the rotation of the lattice with respect to the reference frame (Fig.~\ref{f1}). Given principle 3, $H$ is entirely determined by lattice potential, which will periodically repeat itself. Note that this principle suggests the \textit{least} temporal symmetry of $H$ is discrete, but it does \textit{not} rule out continuous time translation symmetry.

We find the below (isotropic) $H$ respects these general principles and provide a vivid description for CVE. Its outstanding (maybe unique) status is underscored by a series of properties in math and physics. 

\textbf{Hamiltonian and symmetry}. For a Weyl node, we have $H_0^{\lambda}={\lambda}{\hbar}v_F(\boldsymbol{k}{\cdot}\boldsymbol{\sigma})$. By direct sum, one obtains a four-component $H$ of a pair of Weyl nodes.
\begin{equation}
\begin{split}
H=\sum_{\lambda}{\oplus}H_0^{\lambda}+\hat{V}=\sum_{\lambda}{\oplus}{\lambda}{\hbar}v_F(\boldsymbol{k}{\cdot}\boldsymbol{\sigma})-\boldsymbol{\omega}{\cdot}\boldsymbol{L}.\label{eq1}
\end{split}
\end{equation}
Free Weyl is exposed to an axial rotation term $\hat{V}=-\boldsymbol{\omega}{\cdot}\boldsymbol{L}$, which gives energetic preference (the minus sign) for orbital angular momentum $\boldsymbol{L}$ to be aligned with rotation $\boldsymbol{\omega}$. Non-commutation $[H_0^{\lambda},L_j]={\lambda}i\hbar^2(\boldsymbol{k}{\times}\boldsymbol{\sigma})_j$ makes the constructed $H$ non-trivial; on the other hand, the commutation $[H^{\lambda},J_j]=[H^{\lambda},L_j]+[H^{\lambda},s_j]={\lambda}i\hbar^2(\boldsymbol{k}{\times}\boldsymbol{\sigma})_j-{\lambda}i\hbar^2(\boldsymbol{k}{\times}\boldsymbol{\sigma})_j=0$ respects principle 1.

The chiral $H_0^{\lambda}$ (handedness $\lambda$) violates parity symmetry, as Weyl typically does;  $\hat{V}$ breaks the time-reversal symmetry (TRS) as well as the spatial translation symmetry (as $k_x, k_y$ no longer commutate with $H$).

Regarding temporal symmetry, principle 4 ensures a ``bottom line" of discrete $2n\pi/\omega_{\text{l}}$ translations. Next we explain why further combining with principles 1, 3 leads to a time-independent $H$, i.e., $H$ enjoys an enhanced symmetry of continuous $t$ translation. 

Principle 1 stems in a fact: the charge speed along $z$-axis (e.g., CVE) is unaffected by observer's motion in $x$-$y$ plane, such as an axial rotation, thus $j_z$ and the amount of charges transported should be the same for observers with different $\omega_0$. That means we can calculate $j_z$ under a specific reference, as any other should yield the same. Consider an observer sticking to the lattice $\omega_0=\omega_{\text{l}}$ and fermion rotating with $\omega_{\text{f}}=\omega_{\text{l}}+\omega$, i.e., $\omega$ with respect to lattice. Since the lattice potential is static under this frame, principle 3 leads to $\partial_tH=0$. That means CVE can adequately be described by a static quantum $H$.  

Therefore, Eq.~(\ref{eq1}) represents a $H$ under a frame of $\omega_0=\omega_{\text{l}}$. Then, we perform (axial) rotation transformation on $H$, which turns out invariant (up to a phase gauge, see later). That means Eq.~(\ref{eq1}) arises from a specially chosen frame, but also applies to general situations ($\omega_0{\neq}\omega_{\text{l}}$). 

In short, principles 1, 3, 4 together render an enhanced temporal symmetry of $H$, which has been overlooked.

\textbf{Unbounded spectrum}. The Hamiltonian $H$ is non-trivial, as it combines $H_0$ and $\hat{V}$ that are specified by distinct eigen-bases: the plane waves (the continuous limit for Bloch functions) and cylindrical Bessel functions \cite{64}. It extends the semi-classical treatment by (i) including the hole branch -- it is a two-band model; (ii) quantization of the rotation.
\begin{figure}
\centering
\includegraphics[scale=0.44]{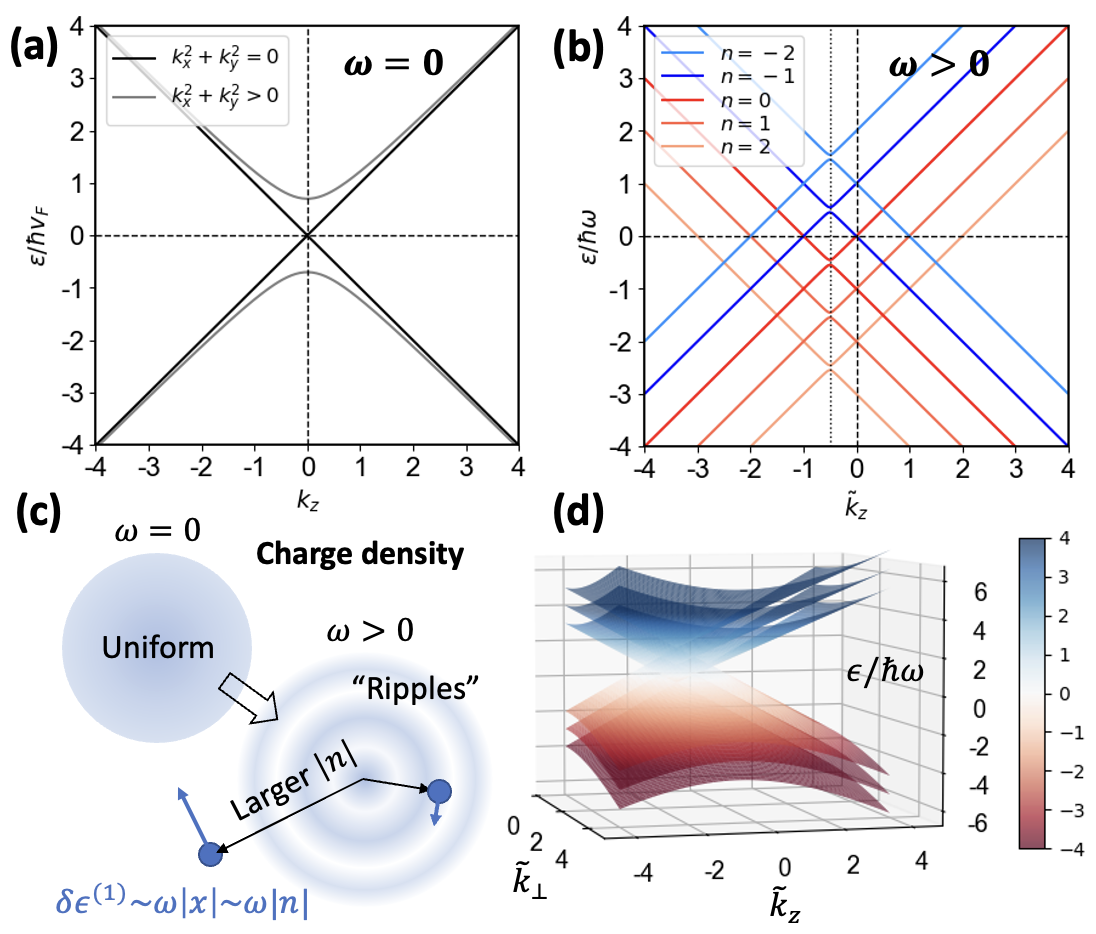}
\caption{\label{fig:epsart2}(color online): Spectra for (a) $H_0$ ($\omega=0$) and (b) $H$ ($\omega>0$). Rotation shifts spectrum along $\tilde{k}_z:=\frac{v_Fk_z}{\omega}$ by $\frac{1}{2}$ (dotted line) and splits a band into infinite ones spaced by $\hbar\omega$ (Floquet features). The infinite $n\in\mathbb{Z}$ (only $|n|\le2$ are plotted) make spectrum unbound. (c) The $|n|$ is in analog with the classical distance $|x|$ to the rotation axis; in real space, charge density ripples form due to disturbance $\omega>0$. (d) Spectrum plotted with full quantum numbers of $H$: $n$, $\tilde{k}_{\perp}:=\frac{v_Fk_{\perp}}{\omega}$, $k_z$. Different branches may intersect. \label{f2}}
\end{figure} 

Without loss of generality, we stipulate $\boldsymbol{\omega}$ along the $z$-axis. Evidently, $H$ should commutate with $J_z$ as requested by principle 1,
\begin{equation}
\begin{split}
[H,J_j]&=[H_0^{\lambda},L_j]+[H_0^{\lambda},s_j]
\\
&={\lambda}{\cdot}i{\hbar}^2(\boldsymbol{k}\times\boldsymbol{\sigma})_j-{\lambda}{\cdot}i{\hbar}^2(\boldsymbol{k}\times\boldsymbol{\sigma})_j=0.\label{eq2}
\end{split}
\end{equation}
Note that $k_z$ commutates with $H$ and $J_z$, but $k_x$, $k_y$ does not commutate with $H$ (broken translation symmetry), nor with $J_z$, $L_z$. Thus, consider a general $[J_z,k_x^n+k_y^n]=\hbar[i{\partial}_{k_x}k_y-i{\partial}_{k_y}k_x,k_x^n+k_y^n]=i{\hbar}(nk_x^{n-1}k_y-nk_xk_y^{n-1})=0$, finding solution $n=2$. After all, the operators mutually commutate: $H$, $J_z$, $k_z$, $k_x^2+k_y^2$, and $\boldsymbol{s}^2$ (but not $\boldsymbol{L}^2$, $\boldsymbol{J}^2$). For common eigenstates, we start with $k_x^2+k_y^2$
\begin{equation}
\begin{split}
k_x^2+k_y^2=-({\partial}_x^2+{\partial}_y^2)=-({\partial}_r^2+\frac{1}{r}{\partial}_r+\frac{1}{r^2}{\partial}_{\phi}^2).\label{eq3}
\end{split}
\end{equation}
The variables $r$ and $\phi$ are separable, leading to common eigenstates with orbital angular momentum $L_z=-i\hbar{\partial}_{\phi}$, whose eigenvalues are $n{\hbar}$. Thus, $-{\hbar}^2{\partial}_{\phi}^2$ can be replaced by $-n^2\hbar^2$, and the equation for eigenstates becomes
\begin{equation}
\begin{split}
-({\partial}_r^2+\frac{1}{r}{\partial}_r+\frac{-n^2}{r^2})\varphi(r)e^{in\phi}=k_{\perp}^2\varphi(r)e^{in\phi}\label{eq4}
\end{split}
\end{equation}
Define variable $\rho:=k_{\perp}r$ with $k_{\perp}{\in}[0,\infty)$, we obtain
\begin{equation}
\begin{split}
(\rho^2{\partial}_{\rho}^2+\rho{\partial}_{\rho}+\rho^2-n^2)\varphi(\rho)=0,\label{eq5}
\end{split}
\end{equation}
which is the Bessel equation, and $n$ is the \textit{order} \cite{64}. In general, the order can be arbitrary $\mathbb{R}$, here since $n$ is quantized by $L_z$, we concern the integers. 

For each $n$, the second-order differential equation has two linearly independent solutions: $J_n(\rho)$ and $Y_n(\rho)$, namely the first and second kinds of Bessel functions. \cite{64} Because $Y_n(0)$ (at the position of rotating axis) is divergent for arbitrary orders, the valid solution is only $J_n(\rho)$. 

The $\varphi(r)e^{in\phi}$ is a common eigenstate of $k_x^2+k_y^2$ and $L_z$, denoted as $|n,k_{\perp}{\rangle}$ (or $|n{\rangle}$ for short); but yet of $H$ or $J_z$. To include them, we examine the off-diagonals of $H$, which are $k_x-ik_y$ and $k_x+ik_y$. We find
\begin{equation}
\begin{split}
[k_x-ik_y,L_z/\hbar]&=[k_x,r_xk_y]+[-ik_y,-r_yk_x]\\
&=[k_x,r_x]k_y+i[k_y,r_y]k_x=k_x-ik_y,\label{eq6}
\end{split}
\end{equation}
where $[r_j,k_l]=i{\cdot}\delta_{j,l}$ has been applied. Define $K^{-}:=k_x-ik_y$, we have $K^-L_z-L_zK^-={\hbar}K^-$, i.e., $K^-L_z-{\hbar}K^-=L_zK^-$. Acting $|n,k_{\perp}{\rangle}$ on both sides,
\begin{equation}
\begin{split}
(K^-L_z-{\hbar}K^-)|n{\rangle}=(n-1){\hbar}K^-|n{\rangle}=L_zK^-|n{\rangle}\label{eq7}
\end{split}
\end{equation}
This suggests $K^-|n,k_{\perp}{\rangle}{\sim}|n-1,k_{\perp}{\rangle}$ (since $[K^-,k_x^2+k_y^2]=0$, $K^-$ does not alter $k_{\perp}$). Similarly, given $K^+:=k_x+ik_y$, we find $K^+|n{\rangle}{\sim}|n+1{\rangle}$. That is,
\begin{equation}
\begin{split}
K^-|n{\rangle}=c_-|n-1{\rangle},~K^+|n{\rangle}=c_+|n+1{\rangle}\label{eq8}
\end{split}
\end{equation}
where $c_{\pm}$ are coefficients to be determined. In view of $(K^-)^{\dagger}=K^+$, we have
\begin{widetext}
\begin{equation}
\begin{split}
&{\langle}n|K^-|n+1{\rangle}={\langle}n|c_-|n{\rangle}=c_-\\
({\langle}n|K^-|n+1{\rangle})^*&={\langle}n+1|K^+|n{\rangle}={\langle}n+1|c_+|n+1{\rangle}=c_+\label{eq9}
\end{split}
\end{equation}
It leads to $c_-=c_+^*$. In addition, $k_x^2+k_y^2=K^-K^+$, thus $c_-c_+=k_{\perp}^2$. We have
\begin{equation}
\begin{split}
|c_+|^2=|c_-|^2=k_{\perp}^2.\label{eq10}
\end{split}
\end{equation}
Thus, $c_-=k_{\perp}{\cdot}e^{-i\delta}$ and $c_+=k_{\perp}{\cdot}e^{i\delta}$ ($c_{\pm}$ are independent of $n$). $K^{\pm}$ connects two neighboring orders of Bessel functions, thus an educated guess for eigen-bases should combine $J_n$ and $J_{n+1}$.
\begin{equation}
\begin{split}
\langle s_z,r,\phi|s,n,k_{\perp}{\rangle}\otimes\langle r_z|k_z\rangle =\begin{pmatrix} b_+^{(s)}(k_{\perp},k_z)J_n(\rho)e^{in\phi} \\ b_-^{(s)}(k_{\perp},k_z)J_{n+1}(\rho)e^{i(n+1)\phi} \end{pmatrix}e^{ik_zr_z},\label{eq11}
\end{split}
\end{equation}
where $s$ refers to the bands, and $b_{\pm}^{(s)}$ are coefficients to be determined. Acting Hamiltonian on the above bases $H|s,n,k_{\perp}{\rangle}$ leads to block diagonalizing
\begin{equation}
\begin{split}
{\lambda}{\hbar}v_F&\begin{pmatrix} b_+^{(s)}k_zJ_ne^{in\phi} & b_-^{(s)}K^-J_{n+1}e^{i(n+1)\phi} \\ b_+^{(s)}K^+J_ne^{in\phi} & -b_-^{(s)}k_zJ_{n+1}e^{i(n+1)\phi} \end{pmatrix}-{\omega}\begin{pmatrix} b_+^{(s)}J_nL_ze^{in\phi}~~,~0 \\ 0,~b_-^{(s)}J_{n+1}L_ze^{i(n+1)\phi} \end{pmatrix} \\
=\hbar &\begin{pmatrix} {\lambda}v_Fk_z-n{\omega} & {\lambda}v_Fk_{\perp}e^{-i\delta} \\ {\lambda}v_Fk_{\perp}e^{i\delta} & -{\lambda}v_Fk_z-(n+1){\omega} \end{pmatrix}\begin{pmatrix} b_+^{(s)}J_ne^{in\phi} \\ b_-^{(s)}J_{n+1}e^{i(n+1)\phi} \end{pmatrix}\label{eq12}
\end{split}
\end{equation}
\end{widetext}
The eigenvalue spectrum ${\epsilon}_{\lambda,s,n}(k_{\perp},k_z)$ turns out to be (independent of $\delta$)
\begin{equation}
\begin{split}
s\hbar\sqrt{(v_Fk_{\perp})^2+({\lambda}v_Fk_z+\frac{\omega}{2})^2}-(n+\frac{1}{2})\hbar\omega,~n{\in}\mathbb{Z},\label{eq13}
\end{split}
\end{equation}
which consists of infinite folds $n$ (Fig.~\ref{f2}(b)), a typical feature for Floquet-type $H$. The spectrum is unbounded due to $n{\in}\mathbb{Z}$. The splitting reflects the additional energy $\delta\epsilon$ brought by rotation. Classically, $\delta\epsilon$ depends on the distance from the rotation axis, i.e., the acceleration or deacceleration effects on a fermion (Fig.~\ref{f2}(c)). The unbounded energy (from both positive and negative sides) is due to the fact that $\delta\epsilon$ tends to infinity as $R\to\infty$. Accordingly, $|n|$ reflects a fermion's classical distance to the axis.  

\subsection{2. Transforming reference frames under Quantum protocol}
It is a misconception to think that $\omega$ in $\epsilon$ (Eq.~(\ref{eq13})) varies with the rotating speed $\omega_0$ of the reference frame. It is also mistaken that the measured energy or distribution necessarily vary with reference frames $\omega_0$, -- a classical impression that is no longer true for quantum. Here, we discuss the quantum protocol and transformation of reference frames within it. 

\textbf{Advance from classical to quantum protocols}. A semi-classical protocol specifies what a configuration is responsible for the chiral transport -- a postulate that bypasses solving the dynamics, but focuses on the state resulting in the currents of Weyl fermions. 

For example, the non-equilibrium $f$ of CVE is determined based on a premise that $f$ appears as equilibrium in a rotating reference frame \cite{8,13,22}. Besides, a protocol for chiral magnetic effect (CME) \cite{47,48} assumes that $f$ is in equilibrium, thereby finding explicit solutions for $f(t)$ is saved; it also assumes $f(\boldsymbol{p},\boldsymbol{x})$ is uniform in real space $\boldsymbol{x}$, which helps remove the currents beyond CME \cite{24}.

Clearly, these protocols (i) were formulated on top of $f$, not $\varphi$; (ii) directly assumed forms of $f$, whose attainability has not been verified by dynamical analysis; (iii) conceptually, relied on equilibrium (or non-equilibrium states convertible to equilibrium).

The quantum protocol (a) transitions from $f$ to a spinful phase-resolved $\varphi$; (b) works with an unbiased wavefunction $\varphi(t)$ to evaluate the form of $f$ instead of assuming $f$ to be (akin to) equilibrium; (c) solves dynamics of microscopic states (standard quantum mechanics) instead of making coarse-averaging arguments (Table~\ref{tab1}).

Concretely, the quantum protocol treats the CVE current as $j_z$ for $\varphi(t)$ that starts from eigenstates of $H_0$ and subsequently evolves under $H_0+\hat{V}$ (Eq.~(\ref{eq1})). At $t=0$, one finds $j_z=0$; upon switching on $\hat{V}$ at $t>0$, $j_z(t){\neq}0$ emerges. The $j_z$ (Sec. 3) is computed by the standard quantum current operator, not involving semi-classical CVE knowledge or expressions \cite{8,18,19,20,20a,21,22,23,24}. 
\begin{table}
\caption{\label{tab:table1} Protocols for CVE: the semi-classical compared with the quantum. \label{tab1}}
%\begin{ruledtabular}
\begin{tabular}{c c c c}
\hline
 & Target & Premise & Sol. dyn. evol. \\
\hline           
\textbf{Semi-classical} & $f(\boldsymbol{p},\boldsymbol{x})$ & Equilibrium $f$ & No \\ 
\textbf{Quantum} & $\varphi$ & Quantum evol. & Yes (solving $\varphi(t)$) \\
\hline
\end{tabular}
%\end{ruledtabular}
\end{table}

To keep the scope focused, we limit our discussion to CVE, which differs from CME or other chiral transports \cite{37,38,39,40,41,42,43,44,45,46,47,48} in roles played by reference frame transformation, non-equilibrium, spatial non-uniformity, etc. 

\textbf{Reference frame transformation}. First, for both classical $\mathcal{C}$ and quantum $\mathcal{Q}$ transformations, the \textit{relative} angular speed $\omega$ is invariant.
\begin{equation}
\begin{split}
\mathcal{C}_{\omega_0\to\omega_0^{'}}[\omega]=\omega,~~\mathcal{Q}_{\omega_0\to\omega_0^{'}}[\omega]=\omega.\label{eq14}
\end{split}
\end{equation}
However, classical energy is generally \textit{not} invariant: $\mathcal{C}[\epsilon]\neq\epsilon$. Because classical energy is associated to non-wave-like particles, and a rotating frame $\omega_0$ definitely alters the measured velocity and force, thereby altering $\epsilon$ \cite{63}. Of peculiar usages are $\epsilon_{\text{lab}}$ and $\epsilon_{\text{rot}}$, which are measured at $\omega_0=0$ and $\omega_0=\omega_{\text{f}}$ (a frame fixed to the fermion) -- clearly, $\epsilon_{\text{lab}}{\neq}\epsilon_{\text{rot}}$.

However, quantum energy (eigenvalue) is associated with waves of the \textit{system}. The quantum $H$ is transformed (between frames $\omega_0$ and $\omega_0^{'}$) by
\begin{equation}
\begin{split}
H{\rightarrow}e^{-iJ_z\delta/\hbar}He^{iJ_z\delta/\hbar},\label{eq15}
\end{split}
\end{equation}
where $\delta=(\omega_0'-\omega_0)t$. Since $[H,J_z]=0$, the $H$ is invariant, except for a phase flexibility $\delta$ in off-diagonals of $H$ (Eq.~(\ref{eq12})) due to $K^+|n{\rangle}=c_+|n+1{\rangle}{\rightarrow}K^+e^{in\delta}|n{\rangle}=c_+e^{i(n+1)\delta}|n+1{\rangle}$. That is, rotation shifts $c_+{\rightarrow}e^{i\delta}c_+$ ($c_-{\rightarrow}e^{-i\delta}c_-$), because $|n{\rangle}$ and $|n+1{\rangle}$ will mismatch the phase under rotation. 

Indeed, the eigenvalue (Eq.~(\ref{eq13})) is independent of $\delta$. Thus, we have 
\begin{equation}
\begin{split}
\mathcal{Q}_{\omega_0\to\omega_0^{'}}[H]=H,~~\mathcal{Q}_{\omega_0\to\omega_0^{'}}[\epsilon]=\epsilon.\label{eq16}
\end{split}
\end{equation}
Such invariance is rendered by the cylindrical symmetry of $H$ -- physically, this means if the observers only rotate themselves ($\omega_0$) and do nothing on the system, the spectrum remains the same. 

The distribution is more subtle. Firstly, a classical $f$ is a function of $(\boldsymbol{p},\boldsymbol{x})$, while quantum $\hat{f}$ is an operator. Secondly, a classical $f$ is not invariant with non-inertial frames $f':=\mathcal{C}[f]{\neq}f$, such as \cite{22}
\begin{equation}
\begin{split}
f'(\epsilon)=f(\epsilon-\boldsymbol{p}{\cdot}(\boldsymbol{w}{\times}\boldsymbol{x})-{\lambda}\hat{\boldsymbol{p}}{\cdot}\boldsymbol{w}/2).\label{eq17}
\end{split}
\end{equation}
This variance has lead to the conjecture: while $f(\epsilon_{\text{lab}})$ is non-equilibrium, $f(\epsilon_{\text{rot}})$ is equilibrium. 
\begin{table}
\caption{\label{tab:table2} Invariance of reference frame transformations under $\mathcal{C}$ and $\mathcal{Q}$ protocols. $j_z$ is required to be ``yes" for $\mathcal{C}$ and $\mathcal{Q}$. But it remains unproved if semi-classical $f$ gives invariant $j_z$ in all frames $\omega_0$. \label{tab2}}
%\begin{ruledtabular}
\begin{tabular}{c c c c c}
\hline
 & $\omega$ & $\epsilon$ & $f$ or $\hat{f}$ & $j_z$ \\
\hline           
\textbf{Classical} & yes & no ($\epsilon_{\text{lab}}{\neq}\epsilon_{\text{rot}}$) & no ($f{\neq}f'$) & unproved \\ 
\textbf{Quant.} & yes & yes ($\epsilon_{\text{lab}}=\epsilon_{\text{rot}}$) & yes ($\hat{f}_{\text{CVE}}=\hat{f}'_{\text{CVE}}$) & yes \\
\hline
\end{tabular}
%\end{ruledtabular}
\end{table}

For quantum $\hat{f}$, transformation is $\hat{f}{\rightarrow}e^{-iJ_z\delta}\hat{f}e^{iJ_z\delta}$ (similar to Eq.~(\ref{eq15})). There is no generic invariance for $\hat{f}$, as it depends on specific states. But the $\varphi(t)$ responsible for CVE is a common eigenstate of $J_z$, thus,
\begin{equation}
\begin{split}
\mathcal{C}_{\omega_0\to\omega_0^{'}}[f]{\neq}f,~~\mathcal{Q}_{\omega_0\to\omega_0^{'}}[\hat{f}_{\text{CVE}}]=\hat{f}_{\text{CVE}}.\label{eq18}
\end{split}
\end{equation}

A physical \textit{requirement} is (for $\mathcal{C}$ and $\mathcal{Q}$) the measured $j_z$ (motions along $z$, not $x$ or $y$) should be identical in all rotating frames $\omega_0$, i.e., $\mathcal{C}[j_z]=j_z$ and $\mathcal{Q}[j_z]=j_z$. Semiclassical $j_z$ depends on $f$, energy, etc., which lack invariance (Table~\ref{tab2}). Although this frame dependence does not rule out an invariant $j_z$, it leaves the issue unsolved. By contrast, in quantum protocol, using $\epsilon_{\text{lab}}$ or $\epsilon_{\text{rot}}$ poses no concern; furthermore, the invariance of quantum protocol (Table~\ref{tab2}) guarantees the invariance of $j_z$. 

\subsection{3. Observables in Quantum formulation}
\begin{figure}
\centering
\includegraphics[scale=0.54]{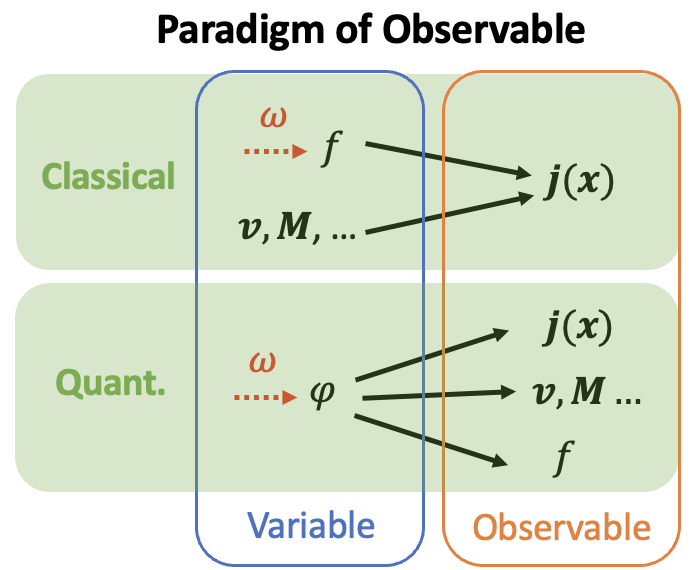}
\caption{\label{fig:epsart}(color online): Causality under classical and quantum protocols, where the $\omega$-dependence enters via $f$ or $\varphi$, respectively. In quantum theory, variables $f$, $\boldsymbol{v}$, $\boldsymbol{M}$, etc. become observables that are mutually hinged by a common $\varphi$. \label{f3}}
\end{figure}

The quantum theory is more than adding ``correction" terms. It changes the paradigm (Fig.~\ref{f3}) by reshaping the causality. It will also supply the valid conditions of semi-classical arguments, such as 2/3 of CVE $j_z$ being contributed by ${\nabla}{\times}\boldsymbol{M}$ \cite{20,22,53a}, which concerns two observables: the current $\boldsymbol{j}$ and the magnetization $\boldsymbol{M}$.

\textbf{Paradigm}. Unlike the non-locality of quantum, the classical paradigm evaluates an observable by a \textit{local} function $o(\boldsymbol{p},\boldsymbol{x})$ ($\boldsymbol{p}, \boldsymbol{x}$ encode full information of a classical state) \cite{49,50,51}. The expectation (for a state $f$) is
\begin{equation}
\begin{split}
\bar{o}=C{\int}o(\boldsymbol{p},\boldsymbol{x})f(\boldsymbol{p},\boldsymbol{x};t)d\boldsymbol{p}{\cdot}d\boldsymbol{x},\label{eq19}
\end{split}
\end{equation}
where $f$ is typically set dimensionless, and $C$ adjusts dimensionality or normalization. To find observable densities in real or momentum spaces, one performs $\int_{\boldsymbol{p}}$ (or $\int_{\boldsymbol{x}}$) on observable function $o(\boldsymbol{p},\boldsymbol{x})$ to yield $\bar{o}(\boldsymbol{x})$ (or $\bar{o}(\boldsymbol{p})$).

The semi-classical theory is based on this paradigm. For example, the CVE current can be derived by an observable function (of Lorentz symmetry) \cite{22}
\begin{equation}
\begin{split}
\boldsymbol{j}(\boldsymbol{x})=-e{\int}\frac{d{\boldsymbol{p}}}{(2\pi{\hbar})^3}(\boldsymbol{v}f(\boldsymbol{p},\boldsymbol{x}))+{\nabla}_x{\times}\boldsymbol{M}(\boldsymbol{x}).\label{eq20}
\end{split}
\end{equation}
In this case, the density $\bar{o}(\boldsymbol{x})$ refers to $\boldsymbol{j}(\boldsymbol{x})$, and observable function $o(\boldsymbol{p},\boldsymbol{x})$ refers to $\boldsymbol{v}(\boldsymbol{p,x})$ and ${\nabla}{\times}\boldsymbol{M}(\boldsymbol{p},\boldsymbol{x})$, local fields in the phase space. 

Conceptually, $\boldsymbol{j}$ is interpreted as the ``consequence", and the fields $f$, $\boldsymbol{v}$, $\boldsymbol{M}$ are interpreted as the ``causes" of the consequence (Fig.~\ref{f3}) -- a semi-classical current contains $f$, $\boldsymbol{v}$, $\boldsymbol{M}$ (possibly others) as independent variables or external fields.

However, quantum offers a paradigm that $\boldsymbol{j}$, $\boldsymbol{M}$, $f$ (even the particle number $N$ which sets the dimension of classical phase space) serve as observables of equal status (Fig.~\ref{f3}) -- the \textit{sole} fundamental variable is the wavefunction $\varphi$, on which all expectation values depend.

Consequently, relationships that classically appear as “one drives the other” become, in the quantum description, correlations between the expectation values of observables jointly determined by a $\varphi$ (or an $\hat{f}$). 

By studying the correlation between ${\langle}j_z{\rangle}$, ${\langle}\boldsymbol{M}{\rangle}$, $f$, etc., one is able to answer questions: (i) the valid condition of semi-classical formula (e.g., Eq.~(\ref{eq20})), i.e., at which condition the correlation (among fields) suggested by semi-classical will hold; (ii) the microscopic nature of observables, such as if $\boldsymbol{M}$ is associated with specific bands or degrees of freedom; (iii) the roles played by non-equilibrium by examining $f$.

\textbf{Observable 1: Current operator}. The current \cite{27,53a} is evaluated by
\begin{equation}
\begin{split}
\boldsymbol{j}(t)={\langle}\frac{{\partial}H(t)}{\partial\boldsymbol{k}}{\rangle}=\text{Tr}[\hat{f}\partial_{\boldsymbol{k}}H(t)].\label{eq21}
\end{split}
\end{equation}
The formula is generic, no restriction to a specific system like Weyl fermion nor a phenomenon like CVE; it is unbiased, without presumption of depending on quantities such as $\boldsymbol{M}$ \cite{20,53a}, Berry curvature \cite{24,45}. It has the wavefunction as the sole input variable, removing the multiple input fields in semi-classical formulas.  

Note that the operator Eq.~(\ref{eq21}) is for the intra-band current, not including the inter-band mechanism, such as \textit{shift currents} \cite{62,65}. This is because only small $\omega$ is concerned about. (Later, we make precise $\omega$ being ``small"). %However, intra-band transport is meaningful when the inter-band transition occurs for $\varphi$ [], known as the \textit{injection current}. 
Nonetheless, the quantum current formula Eq.~(\ref{eq21}) is introduced to play counterpart roles as Eq.~(\ref{eq20}). 

\textbf{Observable 2: magnetization}. Another observable, particularly interested in by this work, is the curl of orbital magnetization: ${\nabla}_x{\times}\boldsymbol{M}$ \cite{20,22,66,67,68,69}. A way of calculating $\boldsymbol{M}$ from microscopic band theory \cite{70,71} is (SI unit)
\begin{equation}
\begin{split}
\boldsymbol{M}&=\boldsymbol{M}_{\text{LC}}+\boldsymbol{M}_{\text{IC}}\\
&=\frac{1}{2}\text{Im}\frac{e}{\hbar}{\int}\frac{d\boldsymbol{k}}{(2\pi)^3}{\langle}{\partial}_{\boldsymbol{k}}u_{\boldsymbol{k}}|{\times}(H_{\boldsymbol{k}}+{\epsilon}_{\boldsymbol{k}})|{\partial}_{\boldsymbol{k}}u_{\boldsymbol{k}}{\rangle},\label{eq22}
\end{split}
\end{equation}
where $u$ represents the periodic part associated with a specific band (thus, Eq.~(\ref{eq22}) is one band contribution). It can be divided into $\boldsymbol{M}_{\text{LC}}$ and $\boldsymbol{M}_{\text{IC}}$. 
\begin{equation}
\begin{split}
\boldsymbol{M}_{\text{LC}}&=\frac{1}{2}\text{Im}\frac{e}{\hbar}{\int}_{BZ}\frac{d\boldsymbol{k}}{(2\pi)^3}{\langle}{\partial}_{\boldsymbol{k}}u_{\boldsymbol{k}}|{\times}H_{\boldsymbol{k}}|{\partial}_{\boldsymbol{k}}u_{\boldsymbol{k}}{\rangle},\\
\boldsymbol{M}_{\text{IC}}&=\frac{1}{2}\frac{e}{\hbar}{\int}_{BZ}-\frac{d\boldsymbol{k}}{(2\pi)^3}{\epsilon}_{\boldsymbol{k}}\boldsymbol{\Omega}_{\boldsymbol{k}}\label{eq23}
\end{split}
\end{equation}
where $\epsilon_{\boldsymbol{k}}$ is band energy and $\boldsymbol{\Omega}$ is berry curvature of that band. $\boldsymbol{M}_{\text{LC}}$ is ``localized" contribution, as it arises from Wannier function of each site (thus replica for every unit cell); $\boldsymbol{M}_{\text{IC}}$ is itinerant contribution, locating mainly on the boundary of bulk, but its magnitude is comparable to $\boldsymbol{M}_{\text{LC}}$ \cite{70}. 

However, Eqs.~(\ref{eq22})-(\ref{eq23}) provide an expectation value $\boldsymbol{M}$ instead of a field $\boldsymbol{M}(\boldsymbol{x})$ \cite{20,60a}. Indeed, $[\boldsymbol{M},\boldsymbol{r}]{\neq}0$ forbids a precise $\boldsymbol{M}(\boldsymbol{x})$ in quantum framework. Moreover, the $\nabla_x$ that appears in Eq.~(\ref{eq20}) is ``coarse-grained" (similar to the $\nabla_x$ that appears in Maxwell equations), different from the fine-resolution $\nabla$ in quantum momentum operator. Thus, we shall later find ways in calculating $\nabla_{x}{\times}\boldsymbol{M}(\boldsymbol{x})$ from the microscopic model.

\section{III. Derive Semi-classical results from Quantum}
The quantum formulation (Sec. 2) endorses several well-known classical results, as discussed in this section. At the same time, quantum sheds light on their validity and offers further insight (Sec. 4).  

\subsection{1. Chiral current under rotation}
A key result from kinetic theory is that, for isotropic Weyl fermion, the CVE current is $j_{\text{CVE}}=\frac{\omega}{(2\pi)^2}{\int}_0^{\infty}2f{\cdot}{\epsilon}d\epsilon$, given $v_F=\hbar=1$ \cite{1,20,21,22,23} -- the basic connotation of CVE. The $j_{\text{CVE}}$ was expressed in terms of a spinless (or spin-averaged) distribution $f(\boldsymbol{p},\boldsymbol{x})$ in the phase space ${\lbrace}\boldsymbol{p},\boldsymbol{x}{\rbrace}$. With the microscopic quantum theory, we should reproduce this from a spinful wavefunctions and a quantized rotation (by contrast, the rotation was described by classical Coriolis forces).

We study the current under the disturbance $\hat{V}=-\boldsymbol{\omega}{\cdot}\boldsymbol{L}$, and the $\hat{f}={\sum}|a_{{\lambda},s,\boldsymbol{k}}|^2|{\psi_{\lambda,s,\boldsymbol{k}}}{\rangle}{\langle}\psi_{\lambda,s,\boldsymbol{k}}|$, where $\lambda$, $s$ stand for the handedness and bands. The CVE current
\begin{equation}
\begin{split}
j_z&=\frac{-e}{(2\pi)^3}\text{Tr}{\int}\hat{f}\frac{1}{\hbar}{\partial}_{k_z}H(\boldsymbol{k})d\boldsymbol{k}\\
&=\frac{-e}{(2\pi)^3}{\sum}_{\lambda,s}{\int}|a_{\lambda,s,\boldsymbol{k}}|^2\frac{1}{\hbar}{\partial}_{k_z}{\epsilon}_{\lambda,s}(\boldsymbol{k})d\boldsymbol{k}.\label{eq24}
\end{split}
\end{equation}
Plugging in the dispersion Eq.~(\ref{eq13}), we have
\begin{equation}
\begin{split}
{\partial}_{k_z}{\epsilon}_{\lambda,s}(\boldsymbol{k})=s{\lambda}v_F{\hbar}\frac{v_Fk_z+\frac{\omega}{2}}{\sqrt{(v_Fk_{\perp})^2+(v_Fk_z+\frac{\omega}{2})^2}}.\label{eq25}
\end{split}
\end{equation}
Then, the (charge) current density splits into two terms: $j_z=j_{\text{ch}}+j_{\text{reg}}$.
\begin{equation}
\begin{split}
j_{\text{ch}}=&\frac{-e}{(2\pi)^3}{\sum}_{\lambda,s}{\int}s|a_{\lambda,s,\boldsymbol{k}}|^2\frac{\lambda v_F\omega}{2\sqrt{(v_Fk_{\perp})^2+(v_Fk_z+\frac{\omega}{2})^2}}d\boldsymbol{k}\\
j_{\text{reg}}=&\frac{-e}{(2\pi)^3}{\sum}_{\lambda,s}{\int}s|a_{\lambda,s,\boldsymbol{k}}|^2\frac{\lambda v_F^2k_z}{\sqrt{(v_Fk_{\perp})^2+(v_Fk_z+\frac{\omega}{2})^2}}d\boldsymbol{k}\label{eq26}
\end{split}
\end{equation}
The $j_{\text{reg}}$ is the regular term that exists for both chiral and non-chiral transports. It stays finite at $\omega\to0$, and its form $j_{\text{reg}}{\to}{\rho}v_z$ tends to the $z$ component of the velocity vector, with odd symmetry
\begin{equation}
\begin{split}
\boldsymbol{v}={\sum}_j{\partial}_{k_j}{\epsilon}_{\lambda,s}(k_x,k_y,k_z)\hat{e}_j={\sum}_j\frac{\lambda{s}v_Fk_j}{k}\hat{e}_j.\label{eq27}
\end{split}
\end{equation}
Thus, the dominant term (zero-order) will vanish. 

\begin{widetext}
The other term $j_{\text{ch}}$ vanishes for non-chiral fermions, thus termed as \textit{chiral}. The $j_{\text{ch}}{\to}0$ at $\omega=0$, thus, it is a current associated with the non-equilibrium peculiarly inspired by rotation. Unlike $j_{\text{reg}}$, it tends to be even with $\boldsymbol{k}$. Expanding $j_{\text{ch}}$ with $\omega$, we obtain
\begin{equation}
\begin{split}
&j_{\text{ch}}=\frac{-e}{(2\pi)^3}{\int}d\boldsymbol{k}{\sum}_{\lambda,s}\lambda s|a_{\lambda,s,\boldsymbol{k}}|^2[{\omega}{\cdot}\frac{1}{2\sqrt{k_{\perp}^2+k_z^2}}\\
&+\frac{\omega^2}{2!}{\cdot}\frac{-k_z}{2v_F(\sqrt{k_{\perp}^2+k_z^2})^3}+\frac{{\omega}^3}{3!}{\cdot}(\frac{9k_z^2}{8v_F^2(\sqrt{k_{\perp}^2+k_z^2})^5}-\frac{3}{8v_F^2(\sqrt{k_{\perp}^2+k_z^2})^3}){\cdots}].\label{eq28}
\end{split}
\end{equation}
This gives an order-by-order expansion for CVE. But the present purpose is the linear CVE coefficient, thus we focus on the first-order $j_{\text{ch}}^{(1)}$
\begin{equation}
\begin{split}
j_{\text{ch}}^{(1)}={\omega}{\cdot}\frac{-e}{(2\pi)^3}{\int}{\sum}_{\lambda,s}\lambda s|a_{\lambda,s,\boldsymbol{k}}|^2\frac{d\boldsymbol{k}}{2k}={\omega}{\cdot}\frac{-e}{(2\pi)^2}{\int}_0^{\infty}{\sum}_{\lambda,s}\lambda s|a_{\lambda,s,\boldsymbol{k}}|^2k{\cdot}dk,\label{eq29}
\end{split}
\end{equation}
where we have used $k:=|\boldsymbol{k}|$ and $d\boldsymbol{k}=4{\pi}k^2dk$. The branch $s=1$ ($s=-1$) has $\epsilon={\hbar}v_Fk$ ($\epsilon=-{\hbar}v_Fk$), leading to a common $k{\cdot}dk=\frac{1}{\hbar^2v_F^2}{\epsilon}{\cdot}d\epsilon$ for both bands.
\end{widetext}

Express the particle/hole distributions with $a_{s,\boldsymbol{k}}$: $f_e=|a_{s=1}|^2$ and $f_h=1-|a_{s=-1}|^2$. For a chiral fermion (all carriers in one node $\lambda=1$, none in the other: $|a_{\lambda=-1,s=1}|^2=0$ and $|a_{\lambda=-1,s=-1}|^2=1$), Eq.~(\ref{eq29}) becomes 
\begin{equation}
\begin{split}
j_{\text{ch}}^{(1)}&={\omega}{\cdot}\frac{-e}{(2\pi)^2}\frac{1}{\hbar^2v_F^2}{\int}(f_e-(-1)f_h)){\epsilon{\cdot}d\epsilon},\label{eq30}
\end{split}
\end{equation}
where $f_h$ has reversed signs twice due to $s=-1$ and $f_h=1-|a_{s=-1}|^2$. For a symmetric situation $f_e=f_h=f$,
\begin{equation}
\begin{split}
j_{\text{ch}}^{(1)}/(-e/\hbar^2)={\omega}{\cdot}\frac{1}{(2\pi)^2v_F^2}{\int}_0^{\infty}2f{\cdot}{\epsilon}{\cdot}d\epsilon.\label{eq31}
\end{split}
\end{equation}
This formula was previously obtained by a semi-classical theory that is aimed to restore Lorentz invariance \cite{20,22,31}. In Sec. 4.3, we provide a more rigorous derivation for it and reveal its valid condition; here, we should point out that $f_e=f_h=f$ applies to $k_BT \gtrsim \mu$, aligned with the wisdom that semi-classical regimes are \textit{not} for the lowest temperature \cite{27}.

On the other hand, at $T\to 0$, it is convenient to express $f$ in chemical potential $\mu$. A Fermi step function was adopted by \cite{53a}, then it yields the same result ($\lambda=1$)
\begin{equation}
\begin{split}
j_{\text{ch}}^{(1)}/(-e/\hbar^2)&={\omega}{\cdot}\frac{1}{(2\pi)^2v_F^2}{\int}_0^{\mu}2{\cdot}{\epsilon}{\cdot}d\epsilon=\alpha~{\omega}{\cdot}(\frac{\mu}{2{\pi}v_F})^2.\label{eq32}
\end{split}
\end{equation}
The dependence on $\mu^2$ is evidently distinct from CME. Physically, it means, if Lifshitz phase transition happens, the CME will reverse the sign \cite{38,46,48}, while the carrier charge type does not affect CVE. 
% Dimensionality analysis: j/(-e)=[w][\mu]^2/([v]^2*[h]^2)=[w][\mu]^2/([\epsilon]^2/[k]^2)=[w][k]^2=[t]^-1[l]^-2, which is the current density (3D)'s dimensionality. Confirmed. 

However, the semi-classical formula is based on relatively higher $T$, thus $f_e=f_h$ might not be satisfied at $T=0$. In that case, suppose $\mu>0$, we have $f_h=0$, and Eqs~(\ref{eq30}), (\ref{eq31}) give $j_{\text{ch}}^{(1)}=\frac{1}{2}{\omega}{\cdot}(\frac{\mu}{2{\pi}v_F})^2$ ($\alpha=\frac{1}{2}$). Indeed, as pointed out by \cite{20}, the CVE is proportional to $(\mu_++\mu_-)(\mu_+-\mu_-)$ with an undetermined overall factor. To connect their results (for free fermion model of a single branch) with the present, we merely need to set $\mu_+=\mu$ and $\mu_-=0$.

Note that $j_{\text{CVE}}$ vanishes at $\mu=0$. This is because $\mu=0$ corresponds to zero carrier density (at $T=0$). That is, Fermi level passes the Weyl node, and the lower (upper) band becomes full (empty) -- become an insulator. Thus, it crucially depends on \textit{metallic} bands.

\subsection{2. Magnetization contribution.}
Another semi-classical result to deduce is that a magnetic contribution $\nabla_x{\times}\boldsymbol{M}$ amounts to 2/3 of the CVE current \cite{22}.

To be concrete, Eq.~(\ref{eq20}) suggests the spatial curl of $\boldsymbol{M}(\boldsymbol{x})$ of Weyl fermion will contribute in addition to the Liouville current; they together form the axial current $j_{\text{CVE}}$, and $\nabla_x{\times}\boldsymbol{M}$ is termed \textit{magnetization contribution} \cite{20,22,53a}, which is 2/3, and the momentum term is 1/3. 

The ${\nabla}_x{\times}\boldsymbol{M}(\boldsymbol{x})$ arises from a magnetic moment coupling $\epsilon=\epsilon_0-\boldsymbol{M}{\cdot}\boldsymbol{B}$ in Lagrangian \cite{20,22}, which is postulated but necessary for restoring Lorentz symmetry \cite{20,22,31}. It is top-down reasoning: given Lorentz symmetry is true, the magnetization contribution should exist without referring to microscopic details.

The present deriving is ``bottom-up": based on a microscopic quantum model (Sec. 2), we show ${\nabla}_x{\times}{\langle}\boldsymbol{M}(\boldsymbol{x}){\rangle}$ equal to 2/3 of ${\langle}j_z{\rangle}$ ($j_{\text{ch}}^{(1)}$ in Eq.~(\ref{eq31})). Essentially, it is about the expectation values of two observables hinged, not seeking assistance from Lorentz symmetry.

The challenge is that ${\nabla}_x$ is ``coarse-grained" (distinguished from $\nabla$). Indeed, the argument to test should be a classical limit. We find a simple way of linking the mesoscopic physics to the microscopic model
\begin{equation}
\begin{split}
\hat{V}=-\boldsymbol{\omega}{\cdot}\boldsymbol{L}:=-\boldsymbol{\omega}{\hbar}{\cdot}(\boldsymbol{r}{\times}\boldsymbol{k}){\approx}-\boldsymbol{\omega}{\hbar}{\cdot}(\boldsymbol{x}{\times}\boldsymbol{k}).\label{eq33}
\end{split}
\end{equation}
The position operator $\boldsymbol{r}$ is replaced by a numerical quantity $\boldsymbol{x}$ that stands for a coarse location and is no longer an operator. The Hamiltonian is parameterized as $H(\boldsymbol{k};\boldsymbol{x})$, and $\nabla_x$ acts on parameter $\boldsymbol{x}$.

The $H_x(\boldsymbol{k};\boldsymbol{x})$ can be viewed as a series of $H$ at distinct $\boldsymbol{x}$ ``glued" together. For each $\boldsymbol{x}$, we can evaluate the quantum expectation of $\boldsymbol{M}$, which means the magnetization in a uniform grain at the location. Accordingly, $\nabla_x$ should mean the coarse scale variation. 

As noted in Sec. 2.3, the orbital magnetic momentum
\begin{equation}
\begin{split}
\boldsymbol{M}=\boldsymbol{M}_{\text{LC}}+\boldsymbol{M}_{\text{IC}},\label{eq34}
\end{split}
\end{equation}
where $\boldsymbol{M}_{LC}$ and $\boldsymbol{M}_{IC}$ refer to the local and itinerant circulation contributions, respectively \cite{70,71}.

The $\boldsymbol{M}_{\text{LC}}$ is given by Eq.~(\ref{eq23}), which relies on $|\partial_{\boldsymbol{k}}u{\rangle}$. In the continuous limit, discrete (lattice) symmetry tends to continuous $i{\partial}_{\boldsymbol{x}}$, and Bloch wave $e^{i\boldsymbol{k}}u_{s,\boldsymbol{k}}(\boldsymbol{r})$ tends to plane wave $e^{i\boldsymbol{k}{\cdot}\boldsymbol{r}}$. That is, the periodic part $u$ is constant, leading to vanishing $\boldsymbol{M}_{\text{LC}}$ up to a modulo \cite{70}. Thus, a continuous model suggests $\boldsymbol{M_{\text{LC}}}$ is spatially uniform, and $\nabla_{x}{\times}\boldsymbol{M}_{\text{LC}}=0$.

Physically speaking, $\boldsymbol{M}_{\text{LC}}$ repeats for each unit cell (independent of $H$ details). This can be seen from the Wannier bases \cite{70,72}
\begin{equation}
\begin{split}
\boldsymbol{M}_{LC}&=\frac{-e}{2NV_{\text{cell}}}{\sum}_j^N{\langle}w_{R_j}|(\boldsymbol{r-\bar{r}_{\boldsymbol{R}_j}}){\times}\boldsymbol{v}|w_{R_j}{\rangle}\label{eq35}
\end{split}
\end{equation}
Every unit cell contributes an identical ${\langle}w_{R_j}|(\boldsymbol{r-\bar{r}_{\boldsymbol{R}_j}}){\times}\boldsymbol{v}|w_{R_j}{\rangle}$ at their locations $\boldsymbol{R}_j$. Thus, $\boldsymbol{M}_{\text{LC}}$ approaches to a uniform background in the continuous limit, when the scale is larger than the lattice constant.

Unlike the local $\boldsymbol{M}_{\text{LC}}$, the itinerant $\boldsymbol{M}_{\text{IC}}$ is non-vanishing for $\nabla_x{\times}\boldsymbol{M}$.
\begin{equation}
\begin{split}
\boldsymbol{M}_{\text{IC}}(\boldsymbol{x})=\frac{1}{2}\frac{e}{\hbar}{\int}\frac{-1}{(2\pi)^3}{\sum}_{\lambda,s}|a_{\lambda,s,\boldsymbol{k}}|^2\epsilon_{\lambda,s}(\boldsymbol{k};\boldsymbol{x}){\boldsymbol{\Omega}}_{\lambda,s,\boldsymbol{k}}{\cdot}d\boldsymbol{k}.\label{eq36}
\end{split}
\end{equation}
Note that only the energy due to $\hat{V}$ provides the dependence of $\boldsymbol{x}$,
\begin{equation}
\begin{split}
&{\nabla}_x{\times}\boldsymbol{M}_{\text{IC}}(\boldsymbol{x})\\
&=\frac{e}{2{\hbar}(2\pi)^3}{\int}{\sum}_{\lambda,s}|a_{\lambda,s,\boldsymbol{k}}|^2{\boldsymbol{\Omega}}_{\lambda,s,\boldsymbol{k}}{\times}{\nabla}_x{\epsilon_{\lambda,s}(\boldsymbol{k};\boldsymbol{x})}{\cdot}d\boldsymbol{k}.\label{eq37}
\end{split}
\end{equation}
Near the Weyl node, the curvature is $\boldsymbol{\Omega}_{s,\boldsymbol{k}}=-{\lambda}s\hat{\boldsymbol{k}}/2k^2$ \cite{71a}, and
\begin{equation}
\begin{split}
{\nabla}_x\epsilon_s(\boldsymbol{k};\boldsymbol{x})&=-{\hbar}{\nabla}_x({\boldsymbol{\omega}}{\cdot}(\boldsymbol{x}{\times}\boldsymbol{k}))\\
&=-\hbar{\nabla}_x(\boldsymbol{x}\cdot(\boldsymbol{k}{\times}\boldsymbol{\omega}))=\hbar(\boldsymbol{\omega}{\times}\boldsymbol{k}).\label{eq38}
\end{split}
\end{equation}
Then we have (given $f_e=f_h=f$)
\begin{equation}
\begin{split}
&{\nabla}_x{\times}\boldsymbol{M}_{\text{IC}}\\
&=\frac{e}{2{\hbar}(2\pi)^3}{\int}{\sum}_{\lambda,s}|a_{\lambda,s,\boldsymbol{k}}|^2\frac{-{\lambda}s\hat{\boldsymbol{k}}}{2k^2}{\times}\hbar(\boldsymbol{\omega}\times\boldsymbol{k}){\cdot}d\boldsymbol{k}\\
&=\frac{-e}{(2\pi)^3}{\int}f\frac{1}{k^2}\hat{\boldsymbol{k}}{\times}(\boldsymbol{\omega}\times\boldsymbol{k}){\cdot}d\boldsymbol{k}.\label{eq39}
\end{split}
\end{equation}
The $\hat{\boldsymbol{k}}{\times}(\boldsymbol{\omega}\times\boldsymbol{k})$ will split into two terms
\begin{equation}
\begin{split}
\nabla_x&{\times}\boldsymbol{M_{\text{IC}}}/(-e/\hbar^2)\\
&=\frac{\hbar^2}{(2\pi)^3}{\int}\frac{f}{k^2}((\hat{\boldsymbol{k}}{\cdot}\boldsymbol{k})\boldsymbol{\omega}-(\hat{\boldsymbol{k}}\cdot\boldsymbol{\omega})\boldsymbol{k})d\boldsymbol{k}.\label{eq40}
\end{split}
\end{equation}
Given the stipulation $\boldsymbol{\omega}=\omega~\hat{e}_z$, the first term $(\hat{\boldsymbol{k}}{\cdot}\boldsymbol{k})\boldsymbol{\omega}=k\boldsymbol~\boldsymbol{\omega}$ gives
\begin{equation}
\begin{split}
\hat{e}_z~\omega{\cdot}\frac{\hbar^2}{(2\pi)^3}{\int}_0^{\infty}\frac{f}{k}{\cdot}4{\pi}k^2dk=\hat{e}_z~{\omega}{\cdot}\frac{1}{(2\pi)^2v_F^2}{\int}_0^{\infty}2f{\cdot}\epsilon{\cdot}d\epsilon\label{eq41}
\end{split}
\end{equation}
The second term gives $(\hat{\boldsymbol{k}}{\cdot}\boldsymbol{\omega})\boldsymbol{k}={\omega}k_z(\hat{e}_xk_x+\hat{e}_yk_y+\hat{e}_zk_z)/k$. The odd terms ($x$ and $y$) vanish in integration, and only ${\omega}k_z^2/k$ remains.
\begin{equation}
\begin{split}
\hat{e}_z\frac{\omega\hbar^2}{(2\pi)^3}&{\int}f\frac{-k_z^2}{k^3}{\cdot}2{\pi}k^2\text{sin}{\theta}d{\theta}{\cdot}dk\\
&=\hat{e}_z\frac{\omega\hbar^2}{(2\pi)^2}{\int}2f{\cdot}k{\cdot}dk{\int}_1^{-1}\frac{\text{cos}^2{\theta}}{2}d(\text{cos}{\theta})\\
&=\hat{e}_z(-\frac{1}{3})~{\omega}{\cdot}\frac{1}{(2\pi)^2v_F^2}{\int}_0^{\infty}2f{\cdot}\epsilon{\cdot}d\epsilon.\label{eq42}
\end{split}
\end{equation}
Add Eq.~(\ref{eq41}) with Eq.~(\ref{eq42}), and compare it with ${\langle}j_z{\rangle}$ in Eq.~(\ref{eq31}) (in unit of $-e/\hbar^2$)
\begin{equation}
\begin{split}
{\nabla}_x&{\times}{\langle}\boldsymbol{M}{\rangle}=\frac{2}{3}{\omega}{\cdot}\frac{1}{(2\pi)^2v_F^2}{\int}_0^{\infty}2f{\cdot}\epsilon{\cdot}d\epsilon=\frac{2}{3}{\langle}j_z{\rangle}.\label{eq43}
\end{split}
\end{equation}
Notably, the 2/3 weight for magnetization contribution holds for varying $v_F$, not limited to perfect free fermions ($v_F=1$). Thus, quantum calculation indicates that the semi-classical current formula Eq.~(\ref{eq20}) remains valid with re-normalized band velocities.

\section{IV. Results beyond the semi-classical theory.}
Deriving established results (Sec. 3) validates the quantum formulation. Next, we seek insights beyond the semiclassical argument.

\subsection{1. Uncover ``void states" due to quantization.}
Quantization might qualitatively alter a classical result, such as breaking classical symmetry, namely \textit{quantum anomaly} \cite{6,24,26}. Here, quantization distinguishes $k_{\perp}=0$ from $k_{\perp}>0$, unlike the uniform classical $k$-space. The structure beyond semi-classical pictures will affect dynamics and transports.
\begin{figure}
\centering
\includegraphics[scale=0.54]{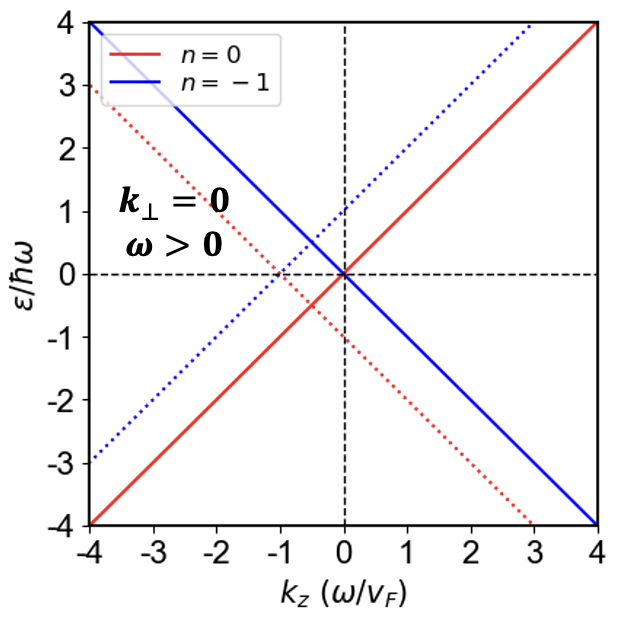}
\caption{\label{fig:epsart}(color online): For $k_{\perp}=0$ eigenstates of $H$, the quantum number $n$ may only take $0$ or $-1$, leading to ${\langle}J_z{\rangle}=\pm1/2$ solely contributed by spin. The void states (non-existing) are dotted lines. \label{f4}}
\end{figure}

\textbf{The case $\boldsymbol{k_{\perp}=0}$}. Reading Eq.~(\ref{eq11}), we find $k_{\perp}=0$ gives a uniform wavefunction (the ripple's wavelength is stretched to infinite),
\begin{equation}
J_n(k_{\perp}r){\equiv}J_n(0)=
\begin{cases}
1, & n=0 \\
0,  & n{\neq}0
\end{cases}\label{eq44}
\end{equation}
Importantly, it leads to ``void states", referring to both components (in Eq.~(\ref{eq11})) vanishing due to $J_n(0)=0$ and $J_{n+1}(0)=0$, simultaneously.
\begin{equation}
\begin{split}
|s,n,k_{\perp},k_z{\rangle}{\sim}\begin{pmatrix} b_+^{(s)}J_n(0)e^{in\phi} \\ b_-^{(s)}J_{n+1}(0)e^{i(n+1)\phi} \end{pmatrix}e^{ik_zr_z}{\equiv}0.\label{eq45}
\end{split}
\end{equation}
This readily happens in view of the Bessel functions: $J_n(0)$ (or $J_{n+1}(0)$) survives only if $n=0$ (or $n=-1$). Consequently, many eigenvalues in Eq.~(\ref{eq13}) are ``void", because their corresponding eigenstates do not exist. 

The occurrence of void states is related to the precondition of block diagonalizing Eq.~(\ref{eq1}) -- the existence of ansatz Eq.~(\ref{eq11}). Otherwise, if the ansatz is null, the eigenvalues Eq.~(\ref{eq13}) become invalid. Seeking the remaining states at $k_{\perp}=0$, we merely need to examine two branches of $|s,0,0,k_z{\rangle}$ and $|s,-1,0,k_z{\rangle}$. 

In addition, $b_{\pm}^{(s)}$ (a function of $k_{\perp}$, $k_z$, independent of $n$, though) also plays a role. For example, $|s,0,0,k_z{\rangle}=(b_+^{(s)},0)^T{\cdot}e^{ik_zr_z}$ (for $\lambda=1$, $s=+1$), where the coefficient $b_+^{(s)}=1$ if $k_z>0$, and $b_+^{(s)}=0$ if $k_z<0$. Thus, the $k_z<0$ half of $|s,0,0,k_z{\rangle}$ is void. Similarly, the $k_z>0$ half of $|s,-1,0,k_z{\rangle}$ is void. After all, the $k_{\perp}=0$ modes are plotted in Fig.~\ref{f4}. 

The void states bring distinctive features for $k_{\perp}=0$ from $k_{\perp}>0$ modes: (i) although the eigenvalues Eq.~(\ref{eq13}) appear asymmetric with $k_z$ (shifted by $v_Fk_z+\omega/2$), the void states make the spectrum regain the symmetry by ``gluing two asymmetric half pieces" (Fig.~\ref{f4}); (ii) the $k_{\perp}=0$ modes do \textit{not} have infinite bands from $J_z$'s quantum number $n$, but only two fragmented bands from $n=0,-1$. 
%(iii) the $k_{\perp}=0$ leads to a spatially uniform state; (iv) robustness to rotation, i.e., all the above features are independent of $\omega$. % Net transport.

The specialty of $k_{\perp}=0$ modes is due to that the rotation symmetry basis (the eigenstate of $H$) must simultaneously respect translation symmetry at $k_{\perp}=0$, thus there is a neat transformation with plane wave bases (the eigenstate of $H_0$). 
\begin{equation}
\begin{split}
|s_z^+,n=0,k_{\perp}=0,k_z{\rangle}&{\sim}|s_z^+,k_{x,y}=0,k_z{\rangle},\\
|s_z^-,n=-1,k_{\perp}=0,k_z{\rangle}&{\sim}|s_z^-,k_{x,y}=0,k_z{\rangle}.\label{eq46}
\end{split}
\end{equation}
We have adopted handedness $\lambda=1$, and the ``$\sim$" means equal up to a phase. Using spin $s_z^{\pm}$ instead of band labels $s$ enables a compact notation; otherwise, segmental definitions have to be used.

\textbf{The case $\boldsymbol{k_{\perp}>0}$}. For $k_{\perp}>0$, the $J_n(k_{\perp}r)$ and $J_{n+1}(k_{\perp}r)$ are spatially dependent and never constantly vanishing, thus no worry for void states. That means, $n$ is not restricted to $0,-1$; for a given $s,n,k_{\perp}$, the $k_z$ takes all values to form a whole band. Different from $k_{\perp}=0$, the $k_{\perp}>0$ display spectrum asymmetric with $k_z$, potentially carrying net velocity. 

Regarding its relationship with plane waves, in principle, we need to perform spatial integration $\int_{\boldsymbol{r}}:={\int}dr_xdr_ydr_z$ between Eq.~(\ref{eq11}) and the below
\begin{equation}
\begin{split}
|s,k_x,k_y,k_z{\rangle}{\sim}\begin{pmatrix} a_+^{(s)}(k_x,k_y,k_z) \\ a_-^{(s)}(k_x,k_y,k_z)\end{pmatrix}e^{i\boldsymbol{k}\cdot\boldsymbol{r}}.\label{eq47}
\end{split}
\end{equation}
However, there lacks a rapid decay for truncation. More importantly, no compatible boundary condition exists between the plane waves and Bessel functions (corresponding to rectangle and cylinder in shapes, respectively). In fact, a direct integration over the entire space ($\int_{\infty}..d\boldsymbol{r}$) suffers from divergence (Appendix A), and an inner product is ill-defined by that. 

Instead, the convergent inner product must rely on symmetry arguments. Derivation is given in Appendix A. Here we quote the result
\begin{equation}
\begin{split}
|s,k_x,k_y,k_z{\rangle}=\sum_{n}^{N_{k_{\perp}}}\frac{e^{-i2\pi(nm)/{N_{k_{\perp}}}}}{\sqrt{N_{k_{\perp}}}}|s,n,k_{\perp},k_z{\rangle},~N_{k_{\perp}}>1,\label{eq48}
\end{split}
\end{equation}
where $N_{k_{\perp}}$ is the degeneracy at $k_{\perp}$, i.e., the number of points $k_x,k_y$ in the ring of $k_{\perp}=\sqrt{k_x^2+k_y^2}$. The integer $m$ is defined by $k_x=k_{\perp}\text{cos}\delta_m$, $k_y=k_{\perp}\text{sin}\delta_m$, where $\delta_m:=\frac{m}{N_{k_{\perp}}}2\pi$ and $m=0, 1...(N_{k_{\perp}}-1)$. The $n$ takes symmetric positive and negative integers, summing zero due to a plane wave carrying ${\langle}L_z{\rangle}=0$.

Note that Eq.~(\ref{eq48}) does \textit{not} apply to $N_{k_{\perp}}=1$ (when the sum involves only one term), and Eq.~(\ref{eq46}) is not a special case of Eq.~(\ref{eq48}), since the quantum number $s, n, k_{\perp}$ are altered by the void states (Fig.~\ref{f4}). 

After all, the Hilbert space spanned by the eigenstates takes a structure
\begin{equation}
\begin{split}
V_\mathcal{H}&\cong(V_{k_{\perp>0}}\oplus V_{k_{\perp}=0})\otimes V_{k_z}\\
&=((V_s\otimes V_n \otimes V_{k_{\perp}>0}) \oplus V_{k_{\perp}=0})\otimes V_{k_z},\label{eq49}
\end{split}
\end{equation}
Therefore, the structure of quantum space proves subtly different from the classical Euclidean ${\lbrace}\boldsymbol{p},\boldsymbol{x}{\rbrace}$ space. 

\subsection{2. Quantum evolution.}
After clarifying the Hilbert space, we solve the quantum evolution and examine such questions: From an initial eigenstate of $H_0$, what a state does it evolve into? What a microscopic wavefunction is responsible for CVE? %If does the semi-classical $f$ coincide with that evaluated by $\varphi(t)$?

\begin{figure}
\centering
\includegraphics[scale=0.57]{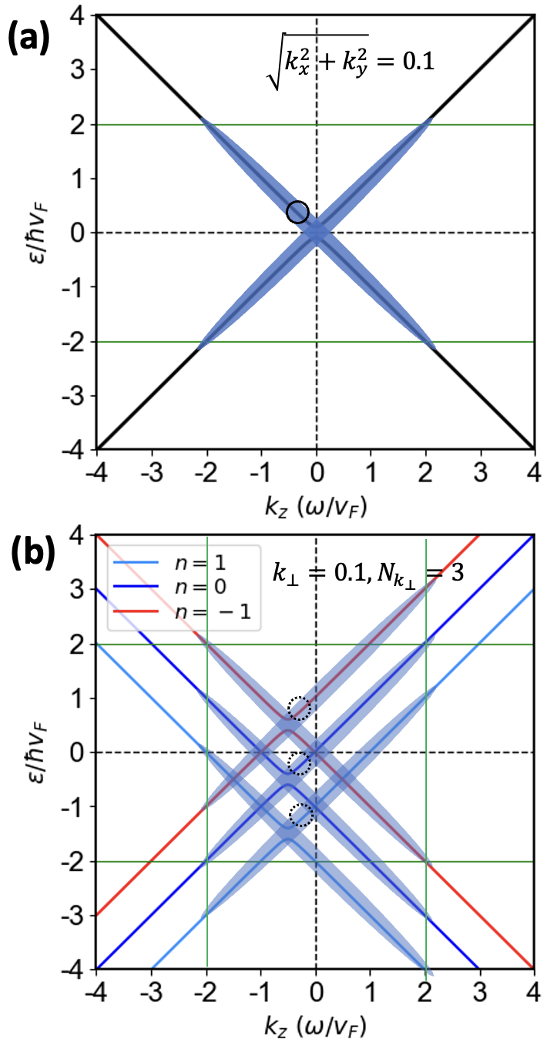}
\caption{\label{fig:epsart}(color online): (a) An initial distribution and (b) its resulting distribution after rotation. The evolution of a single state (for a particular $s$ and $k_z$) is shown in where velocities have reversed (the circles). The $N_{k_{\perp}}=3$ is selected for demonstration; in general, it splits into $N_{k_{\perp}}$ equal-weighted branches. \label{f5}}
\end{figure}

An initial state is characterized by a density density operator that is specified by coefficients $a$ associated with $H_0$'s eigenstates $|s,k_x,k_y,k_z{\rangle}$ (set $\lambda=1$)
\begin{equation}
\begin{split}
\hat{f}(0)=\sum_{\boldsymbol{k}}|a_{s,\boldsymbol{k}}|^2|s, \boldsymbol{k}{\rangle}{\langle}s,\boldsymbol{k}|,\label{eq50}
\end{split}
\end{equation}
where $\boldsymbol{k}:=k_x,k_y,k_z$ for short. The evolution of the density operator is
\begin{equation}
\begin{split}
\hat{f}(t)=\sum_{\boldsymbol{k}}|a_{s,\boldsymbol{k}}|^2U(t)|s, \boldsymbol{k}{\rangle}{\langle}s,\boldsymbol{k}|U^{\dagger}(t),\label{eq51}
\end{split}
\end{equation}
where $U(t)$ is associated with $H=H_0+\hat{V}$. Thus, $U(t)$ can conveniently be solved under the bases $|s,n,k_{\perp},k_z{\rangle}$, taking a diagonal form
\begin{equation}
\begin{split}
U(t)={\sum}_{s,n,k_{\perp},k_z}&\text{exp}(-i{\epsilon_{s,n}(k_{\perp},k_z)t}/{\hbar})\\
&\times|s,n,k_{\perp},k_z{\rangle}{\langle}s,n,k_{\perp},k_z|.\label{eq52}
\end{split}
\end{equation}
The bases for $\hat{f}$ can be switched with Eq.~(\ref{eq48}) For easy handling, we adopt a polar label ${\lbrace}{s,\delta_m,k_{\perp},k_z}{\rbrace}:={\lbrace}{s,\boldsymbol{k}}{\rbrace}$, where $k_x=k_{\perp}\text{cos}\delta_m,~k_y=k_{\perp}\text{sin}\delta_m$ (a mere designation change for convenience later).
\begin{widetext}
\begin{equation}
\begin{split}
&\hat{f}(0)=\sum_{\delta_m,k_{\perp},k_z}|a_{s,\delta_m,k_{\perp},k_z}|^2|s, \delta_m,k_{\perp},k_z{\rangle}{\langle}s,\delta_m,k_{\perp},k_z|\\
&=\sum|a|^2\left(\sum_{n}\frac{e^{-i\frac{2\pi(nm)}{N_{k_{\perp}}}}}{\sqrt{N_{k_{\perp}}}}|s, n,k_{\perp},k_z{\rangle}\right)\left(\sum_{n'}\frac{e^{i\frac{2{\pi}(n'm)}{N_{k_{\perp}}}}}{\sqrt{N_{k_{\perp}}}}{\langle}s,n',k_{\perp},k_z|\right)\\
&=\sum_{\delta_m,k_{\perp},k_z}\sum_{n,n'}\frac{|a_{s,\delta_m,k_{\perp},k_z}|^2}{N_{k_{\perp}}}e^{-i\frac{2m\pi}{N_{k_{\perp}}}(n-n')}|s, n,k_{\perp},k_z{\rangle}{\langle}s,n',k_{\perp},k_z|.\label{eq53}
\end{split}
\end{equation}
\end{widetext}
In general, there exist off-diagonals $n{\neq}n'$, while the occupancies over the eigenstates $|s,n,k_{\perp},k_z{\rangle}$ merely depend on the diagonals. A representative density evolution ($k_{\perp}>0$) is evaluated with Eq.~(\ref{eq52}) and illustrated in Fig.~\ref{f5}. 

A key finding is that a symmetric initial density $\hat{f}(0)$ like Fig.~\ref{f5}(a) should evolve into an $\hat{f}(t)$ \textit{asymmetric} with $k_z$ (Fig.~\ref{f5}(b)), a microscopic state carrying net currents along $z$. Since $[k_z,H]=0$, $k_z$ is conserved in evolution. Thus, it is distinct from a current driven by an electric field, which accelerates $k_z$.

Additionally, the wavefunction becomes an (equal-weighted) superposition of $N_{k_{\perp}}$ folds, where $N_{k_{\perp}}$ is the degeneracy at a particular $k_{\perp}$, i.e., the number of ${\lbrace}k_x,k_y{\rbrace}$ points in a ring of radius $k_{\perp}$. The splitting is symmetric with respect to $n$, making the total $J_z=0$. This is because $J_z$ is conserved ($[J_z,H]=0$).

%where $c_{n,k_{\perp}}:={\langle}s,n,k_{\perp},k_z|s,k_x,k_y,k_z{\rangle}$. 
Physically, the splitting can be understood as the rotation perturbs a spatially uniform state into a series of ``ripple-shaped" Bessel states $|s,n,k_{\perp},k_z{\rangle}$ (Fig.~\ref{f2}(c)) -- the ripple's shape depends on $n$, and the (radial) wavelength is controlled by $k_{\perp}$. 

The result is directly based on the solution of equations of motion, free of additional approximations. It suggests that the microscopic state for CVE is not the lowest energy one -- the lower-energy $n$ bands exist but unattainable in dynamical evolution. This underscores the origin of CVE in non-equilibrium meta-stable states. 

\subsection{3. The role of isotropic approximation.}
Semi-classical results, e.g., $j=\frac{\omega}{(2{\pi}v_F)^2}{\int}f(\epsilon)2{\epsilon}{\cdot}d\epsilon$, rely on isotropy. The influence by anisotropy remains unexplored. Here, we study roles of this approximation in quantum contexts and find that anisotropy is more than adopting an orientation-resolved $f$. 

In Sec. 4.2, we solved $\hat{f}(t)$, with which CVE currents can be solved. In general, $\hat{f}(t)$ in $H$'s eigenbases contains off-diagonals, causing at least two complexities. First, $\dot{f}(t)\neq0$ (time-dependence arises from off-diagonals), thus static distribution $f$ might be unattainable. Second, off-diagonals of observable operators, (e.g., $v_z $ is non-diagonal due to $[v_z,H]{\neq}0$) come into play when $\text{Tr}[\hat{f}v_z]$ is calculated -- this makes slopes of spectrum (procedures in Sec. 3.1) incomplete for $j_z$.

\begin{widetext}
However, $\hat{f}$ can be simplified by (in-plane) isotropic approximation, i.e., the initial state $a_{s,\delta_m,k_{\perp},k_z}$ is independent of angle $\delta_m$. Then, Eq.~(\ref{eq53}) becomes
\begin{equation}
\begin{split}
\hat{f}(0)&=\sum_{k_{\perp},k_z}|a_{s,k_{\perp},k_z}|^2\sum_{n,n'}\sum_{\delta_m}\frac{e^{-i\frac{2m\pi}{N_{k_{\perp}}}(n-n')}}{N_{k_{\perp}}}|s, n,k_{\perp},k_z{\rangle}{\langle}s,n',k_{\perp},k_z|\\
&=\sum_{k_{\perp},k_z}|a_{s,k_{\perp},k_z}|^2\sum_{n,n'}\delta_{n,n'}|s, n,k_{\perp},k_z{\rangle}{\langle}s,n',k_{\perp},k_z|=\sum_{n,k_{\perp},k_z}|a_{s,k_{\perp},k_z}|^2|s, n,k_{\perp},k_z{\rangle}{\langle}s,n,k_{\perp},k_z|.\label{eq54}
\end{split}
\end{equation}
With the in-plane isotropic approximation, the off-diagonals of $\hat{f}$ vanish. Given $\hat{f}(0)$ is diagonal in eigenbases of $H$, the $\hat{f}(t)$ takes forms of
\begin{equation}
\begin{split}
\hat{f}(t)&=\sum_{n,k_{\perp},k_z}|a|^2e^{-i{\epsilon}}|s, n,k_{\perp},k_z{\rangle}{\langle}s,n,k_{\perp},k_z|e^{i{\epsilon}}=\hat{f}(0),\label{eq55}
\end{split}
\end{equation}
\end{widetext}
where the dynamic phases cancel. Thus, isotropic approximation guarantees forming a static $\hat{f}(t)=\hat{f}(0)$ on rotation, and $j_z=\rho\text{Tr}[\hat{f}(t)v_z]$ is readily evaluated by the initial distribution $\hat{f}(0)$ 
\begin{equation}
\begin{split}
j_z(t)&=\frac{\rho}{\hbar}\text{Tr}[\hat{f}(0)\partial_{k_z}H]\\
&=\frac{\rho}{\hbar}\sum_{n,n'}|a|^2{\langle}n'|n{\rangle}{\langle}n|(\partial_{k_z}H)|n'{\rangle}\\
&=\frac{\rho}{\hbar}\sum_n|a|^2{\langle}n|(\partial_{k_z}H)|n{\rangle},\label{eq56}
\end{split}
\end{equation}
where $|n{\rangle}$ stands for $|s,n,k_{\perp},k_z{\rangle}$. Although $|s, n,k_{\perp},k_z{\rangle}$ is simpler for eigenvalues of $H$, handling $\sum_{n}^{\infty}$ and $\sum_{k_{\perp}}$ is cumbersome. On the other hand, for convenient summation, $j_z$ is expressed in $H_0$'s eigenbases $|s,\delta_m,k_{\perp},k_z{\rangle}$
\begin{equation}
\begin{split}
j_z=\frac{\rho}{\hbar}\sum_{\boldsymbol{k}}|a_{s,k_{\perp},k_z}|^2{\langle}\boldsymbol{k}|(\partial_{k_z}H)|\boldsymbol{k}{\rangle},\label{eq57}
\end{split}
\end{equation}
where $|\boldsymbol{k}{\rangle}$ is shorthand for $|s,\delta_m,k_{\perp},k_z{\rangle}$. (In above, we have adopted intra-band contribution, as semi-classical results required  a single-fold $f$ \cite{49,50,51,52}).

The problem of using $|\boldsymbol{k}{\rangle}$ is how to evaluate ${\langle}{\boldsymbol{k}}|({\partial}_{k_z}H)|{\boldsymbol{k}}{\rangle}$, since $|{\boldsymbol{k}}{\rangle}$ is \textit{not} an eigenstate of $H$ or $L_z$. Note that $|{n,k_{\perp},k_z}{\rangle}$ is composed by $|{\boldsymbol{k}}{\rangle}$ living in a ``shell" $I_{k_{\perp}}={\lbrace}(k_x,k_y)|k_x^2+k_y^2=k_{\perp}^2{\rbrace}$. Given $|{\boldsymbol{k}}{\rangle}{\in}I_{k_{\perp}}$, $k_{\perp}{\neq}k_{\perp}'$, we should have ${\langle}{n,k_{\perp}',k_z}|{\boldsymbol{k}}{\rangle}=0$. Then, the superposition becomes $|{\boldsymbol{k}}{\rangle}=\sum_{n}c_n|{n}{\rangle}$. The explicit form of coefficients $c$ are given by Appendix A. Then, the diagonals of current matrix become
\begin{equation}
\begin{split}
&{\langle}{\boldsymbol{k}}|({\partial}_{k_z}H)|{\boldsymbol{k}}{\rangle}=\sum_{n}|c_n|^2{\langle}{n}|({\partial}_{k_z}H)|{n}{\rangle}\\
&=\sum_{n}|c_n|^2\left({\partial}_{k_z}({\langle}{n}|H|{n}{\rangle})-{\langle}\partial_{k_z}n|H|n{\rangle}-{\langle}n|H|\partial_{k_z}n{\rangle}\right),\label{eq58}
\end{split}
\end{equation}
where the brackets $(\partial_{k_z}..)$ indicate the derivative range. The last two terms should vanish as they are equal to $-\epsilon(\langle\partial_{k_z}n|n{\rangle}+{\langle}n|\partial_{k_z}n{\rangle})=-\epsilon\partial_{k_z}({\langle}n|n{\rangle})$. The remaining becomes
\begin{equation}
\begin{split}
{\langle}{\boldsymbol{k}}|({\partial}_{k_z}H)|{\boldsymbol{k}}{\rangle}&={\partial}_{k_z}\epsilon_{n}(k_{\perp},k_z)\sum_{n}|c_n|^2\\
&={\partial}_{k_z}\epsilon_{n}(k_{\perp},k_z),\label{eq59}
\end{split}
\end{equation}
where ${\partial}_{k_z}\epsilon_{n}(k_{\perp},k_z)$ is factored out for $n$-independence of $\epsilon$ (Eq.~(\ref{eq13})). This is due to the features of $H$ (related to Floquet): $n$ gives a series of bands of distinct energies but identical slopes.

Then, Eq.~(\ref{eq57}) becomes Eq.~(\ref{eq24}) after taking the continuous $\sum_{\boldsymbol{k}}{\to}\frac{V}{(2\pi)^3}{\int}d\boldsymbol{k}$ and $\rho=-e/V$. It can be viewed as a more water-proof derivation of Sec. 3.1, which has skipped justifications for (i) the integration being performed with $dk_xdk_y$, which are not good quantum numbers for $H$ (ii) the coefficient $a$ being taken time-independent (iii) the energy $\epsilon$ is for unperturbed $H_0$.

Now it becomes clear: firstly, it is isotropy that vanishes the off-diagonals of $\hat{f}$ under $|\boldsymbol{k}{\rangle}$ bases and makes the variable substitution equivalent; secondly, static $\hat{f}$ is a quantum evolution consequence under isotropy, not an assumption; thirdly, since $f(\epsilon)$ is the initial distribution, the $\epsilon$ is for $H_0$. 

\begin{figure}
\centering
\includegraphics[scale=0.55]{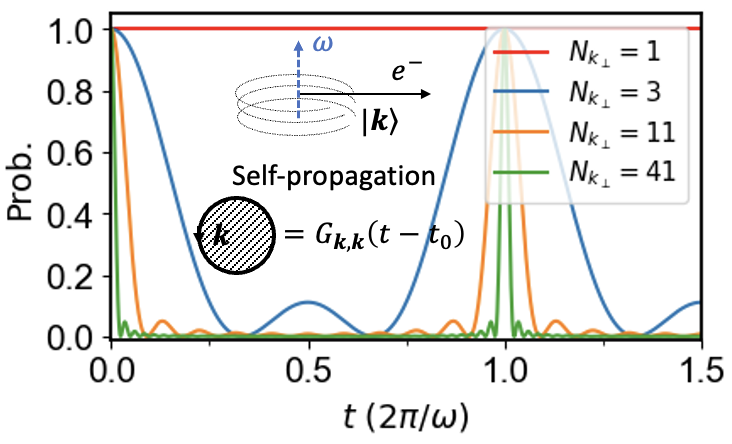}
\caption{\label{fig:epsart}(color online): The time-dependence of occupancy of state $\boldsymbol{k}$, which turns out periodic with $2\pi/\omega$. Large degeneracy $N_{k_{\perp}}$ tends to classical limit, wherein the self-propagation function becomes periodic $\delta$-function. The $N_{k_{\perp}}=1$ (red) corresponds to a single mode $k_{\perp}=0$, which regains isotropic condition and static $f$. \label{f6}}
\end{figure}

On the other hand, anisotropy possibly leads to $\dot{\hat{f}}(t)\neq0$, thus time dependent observables. For example, with an anisotropic initial $\hat{f}(0)=|\boldsymbol{k}{\rangle}{\langle\boldsymbol{k}}|$ (i.e., a form of $\delta$-function $a_{s,\delta_m,k_{\perp},k_z}=\delta_{m,m_0}$), we obtain time-dependent occupancy of $|\boldsymbol{k}{\rangle}$ (Fig.~\ref{f6}). Notably, the semi-classical formula  have uncritically assumed a static $f$ (in time scale $1/\omega$).

\subsection{4. Physical validity and meaning of $f(\epsilon)$ in semi-classical protocols.}
Semi-classical protocols may presume one or multiple of the below: $f(\epsilon)$ takes forms of fermi distribution; $f(\epsilon)$ appears as equilibrium in the rotating (non-inertia) frame, etc. We test the validity with quantum wavefunctions.

\textbf{Fermi distribution works for CVE?} For non-equilibrium phenomena \cite{26,40,51,73}, fermi $f_F(\epsilon)=1/(e^{\beta\epsilon}+1)$ is not for certain; particularly, CVE lacks a ground state (lacks a ``destination" for thermal relaxation) due to its boundless spectrum. 

\begin{figure}
\centering
\includegraphics[scale=0.5]{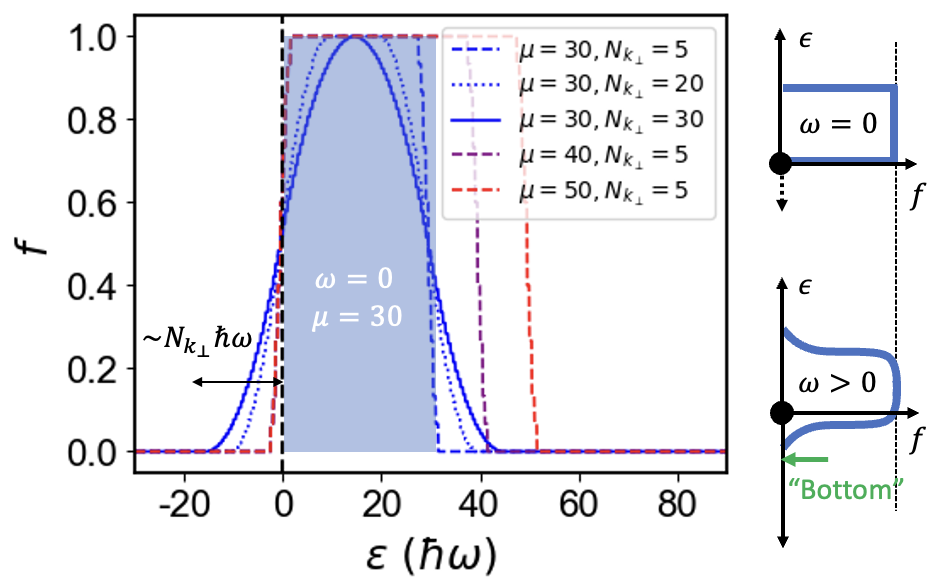}
\caption{\label{fig:epsart}(color online): $f(\epsilon)$ evaluated by Schr\"odinger equation with isotropic $\hat{f}(0)$ at zero temperature, compared with $f_F(\epsilon)$ of $\mu=30$ (shadowed). The semi-positive spectrum is extended to $-\infty$ at $\omega>0$, due to $n\in\mathbb{Z}$. A ``bottom" could be defined at $-N_{k_{\perp}}\hbar\omega$, below which states are available but empty (dynamically unattainable), thus violating $f_F(\epsilon)$. \label{f7}}
\end{figure}

Indeed, Fig.~\ref{f6} already shows that $f$ for anisotropic CVE is time-dependent, unlike the static $f_F$ \cite{17}. Then, does $f_F$ apply to isotropic CVE? Consider an initial isotropic filling from $0$ to chemical potential $\mu$, and solve $\hat{f}(t)$ for $t>0$ (based on Sec. 4.2). The distribution at $T=0$ is shown in Fig.~\ref{f7}. 

In this case, $f(\epsilon)$ is static and broadened owing to the rotation quanta $n$. Consequently, even at $T=0$, fermions spread into higher and lower energies. This is distinct from thermal spreading which occurs at one end (near $\mu$) by $k_BT$; rotation affects from both ends (near both $0$ and $\mu$) by ${\sim}N_{k_{\perp}}\hbar{\omega}$ due to rotation quanta adsorbed or emitted.

We can re-express $N_{k_{\perp}}\approx\frac{k_F}{(2\pi/R)}\cdot2\pi=k_FR$, where $k_F$ is the fermi wave vector. Then, the spreading range becomes $\sim \hbar k_F \omega R$. This provides an intuitive understanding: the spread width arises from the rim speed imposed by rotation $\omega R$ multiplied with the fermi filling. 

Accordingly, $f(\epsilon)$ may take a ``bump" shape. On the other hand, given the limit $\mu{\gg}N_{k_{\perp}}\hbar{\omega}$, at which ``$\hbar{\omega}$" is small and its influence on fermi occupancy (the ``spreading") is negligible, the distribution tends to square-shaped. 

In addition, $f_F(\epsilon)$ suggests the low-energy states are preferentially occupied. However, Fig.~\ref{f7} indicates a ``bottom", below which the states are available and $f(\epsilon)=0$, while $f_F(\epsilon)=1$. Thus, fermi distribution only applies to states well above the bottom energy: $\epsilon{\in}(N_{k_{\perp}}\hbar\omega,\infty)$. 

Thus, the mere low-$T$ is inadequate for $f_F(\epsilon)$ to hold, which underscores CVE's origin in non-equilibrium. It requires two extra conditions: (i) the chemical potential $\mu$ is much larger than rotation quanta: $\mu/\hbar{\omega}{\gg}N_{k_{\perp}}$; (ii) restricted to states well above the bottom $\epsilon\in(N_{k_{\perp}}\hbar\omega,\infty)$. Clearly, as $\omega{\to}0$, we have $\mu/\hbar{\omega}{\to}\infty$ and $(N_{k_{\perp}}\hbar\omega,\infty)\to(0, \infty)$ -- then, $f_F$ applies to the whole range. 
\begin{figure}
\centering
\includegraphics[scale=0.7]{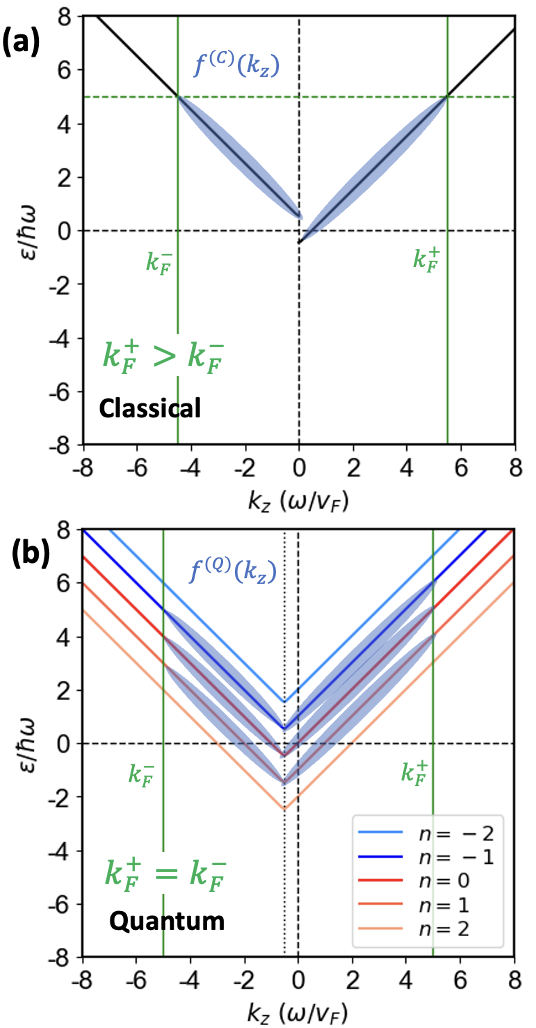}
\caption{\label{fig:epsart}(color online): Distributions over $k_z$ (shadow) at $k_{\perp}\to0$: (a) classical $f^{(C)}$ and (b) quantum $f^{(Q)}$. Classical $f^{(C)}$ is for a small region around $\boldsymbol{x}\to0$, and energy shift is $\hat{\boldsymbol{p}}{\cdot}{\boldsymbol{\omega}}=\hat{\boldsymbol{k}}\hbar\boldsymbol{\omega}$, making a discontinuous spectrum at $k_z=0$. A number of $n$-bands are plotted in (b), but there exist more. \label{f8}}
\end{figure}

\textbf{How well can $f$ approximate equilibrium in rotating frame?} the CVE may involve a premise \cite{22,23} when in-lab non-equilibrium $f(\epsilon)$ is transformed to a co-rotating frame $f'(\epsilon):=\mathcal{C}_{0{\to}\omega}[f(\epsilon)]$, it should appear like equilibrium, i.e., $f'(\epsilon)=f_F(\epsilon)$. This is conditionally true (Fig.~\ref{f7}). Next, we test the equilibrium assumption from broader aspects, e.g., $f(k_z)$.

Acting frame transformation $\mathcal{C}_{\omega\to0}$ on both sides of $f'(\epsilon):=\mathcal{C}_{0{\to}\omega}[f(\epsilon)]$, one obtains $f(\epsilon)=\mathcal{C}_{\omega\to0}[f'(\epsilon)]$. Combining with $f'(\epsilon)=f_F(\epsilon)$, we have
\begin{equation}
\begin{split}
f(\epsilon)=f_F(\epsilon'),\label{eq60}
\end{split}
\end{equation}
where $\epsilon$ and $\epsilon'$ are the (classical) energies measured in the lab and co-rotating reference frames \cite{22}. Equation~(\ref{eq60}) links non-equilibrium $f(\epsilon)$ to a simpler fermi $f_F(\epsilon')$, where $\epsilon'$ is given by Eq.~(\ref{eq17}) -- This yields approximate $f(\epsilon)$ in effort-saving way, rather via solving equations of motions. With $f(\epsilon)$, one can derive distributions over $k_z$ (Fig.~\ref{f8}(a)).

For quantum, the frame transformation is specified by Eq. To be specific, the density operator in the co-rotating frame is
\begin{equation}
\begin{split}
\hat{f}'=e^{-iJ_z({\omega}t)}\hat{f}e^{iJ_z({\omega}t)}.\label{eq61}
\end{split}
\end{equation}
The distribution $f(k_z)$ yielded from quantum equations of motion is plotted in Fig.~\ref{f8}(b).  

The semi-classical $f^{(C)}$ Fig.~\ref{f8}(a) which is based on equilibrium protocols) resembles the quantum $f^{(Q)}$ Fig.~\ref{f8}(b), given splittings by $\hbar\omega$ are small; moreover, $f^{(C)}$ and $f^{(Q)}$ derive an identical $j_z$. On the other hand, $k_z$ is conserved in $f^{(Q)}$, thus filling is symmetric $k_F^+=k_F^-$, and the net current is due to the spectrum shift by $\omega/2v_F$ (dotted in Fig.~\ref{f8}(b)); while $k_z$ is not conserved in $f^{(C)}$, because the filling is up to a common chemical potential due to the equilibrium protocol, leading to $k_F^+>k_F^-$.%(given $\chi=1$, $\omega>0$). 

Thus, assuming equilibrium in a co-rotating frame gives a reasonable $f^{(C)}(\epsilon)$ (with non-quantized $\epsilon$) when $\mu/\hbar\omega\gg N_{k_{\perp}}$, but leads to an $f^{(C)}(k_z)$ that qualitatively differs from $f^{(Q)}(k_z)$. Nonetheless, the approximate $f^{(C)}$ leads to $j_z$ that coincides with quantum theory. This is because a common $j_z$ can be yielded from multiple distributions. The semi-classical $f$ can be viewed as an ``effective" distribution (in terms of velocity slopes) for charge current $j_z$; however, its validity cannot be extended to every other observable. For example, although missing the splitting sub-bands of $n$ (Fig.~\ref{f8}(b)) has no effect on currents of charge, it can cause deviations in observables related to spin.  

\textbf{Physical meaning of semi-classical distribution.} After all, we are able to clarify the connotation of $f$ functions (not an operator) that appear in semi-classical. First, the $f$ contained in $j=\frac{\omega}{(2{\pi}v_F)^2}{\int}f2{\epsilon}{\cdot}d\epsilon$ must be \textit{isotropic} (but \textit{not} necessarily equilibrium); if $f$ is anisotropic, the expression of current becomes invalid (for example, it fails to capture the time dependence, see Fig.~\ref{f6}). Second, the semi-classical $f(\epsilon)$ is derived from coefficients in density operator $\hat{f}(0)$ at $t=0$ (Eq.~(\ref{eq54})), thus it stands for the \textit{initial} distribution $f(\epsilon;t=0)$, rather than the instantaneous nor that has been modified by rotation $f(\epsilon;t>0)$ (e.g., it corresponds to the shadowed in Fig.~\ref{f7}, rather than those distorted lines). Third, $f(\epsilon)$ arises from $|a|^2$, thus it means the filling of a state (values $\in[0,1]$ for fermion), rather than the density of states.

\subsection{5. Quantum regimes for CVE: Two scales uncovered.}
Physical phenomena typically involve scales in energy, length, etc. that dictate classical-quantum crossover. Quantum formulation (defined in Sec. 2) underscores two scales, which might have been overlooked by semi-classical treatments. 

The first concerns the ratio of two velocities: the rotating speed at the rim of the system and the fermi velocity $v_F$. The present treatment relies on
\begin{equation}
\begin{split}
\omega R/v_F \ll 1.\label{eq62}
\end{split}
\end{equation}

Intuitively, $\omega R/v_F\ll1$ means that the maximum rotation velocity (at the rim) is small compared with the intrinsic $v_F$. This condition corresponds to a situation of negligible inter-band transition (for the two bands defined by $H_0$ in Eq.~\ref{eq1}) in quantum context.  
\begin{figure}
\centering
\includegraphics[scale=0.45]{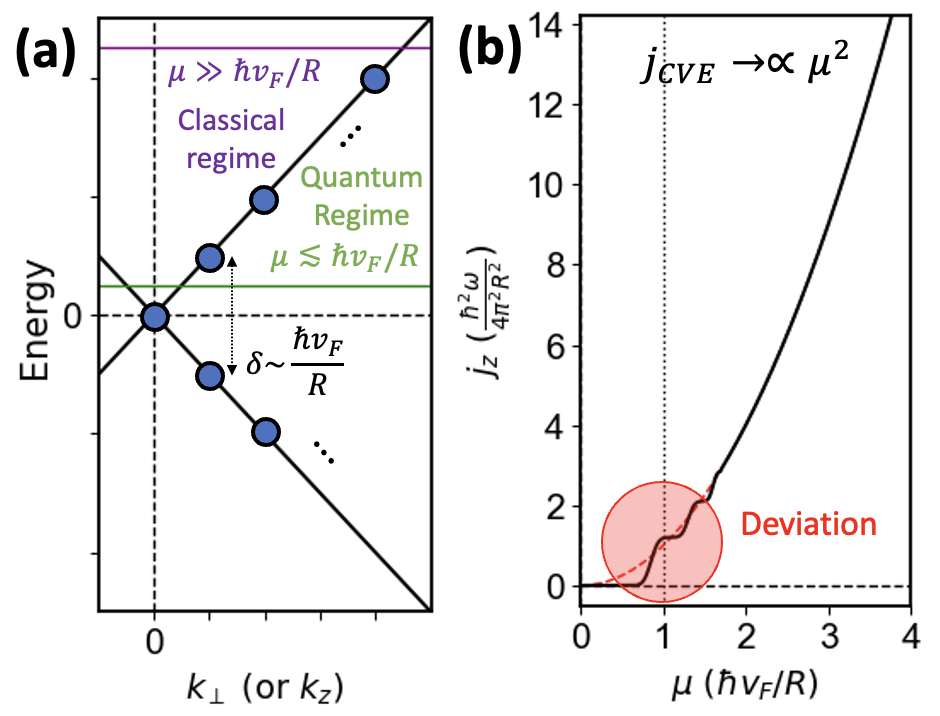}
\caption{\label{fig:epsart}(color online): (a) Quantum or classical regimes tuned by $\mu/(\hbar v_F/R)$. Values of $k$ (solid dots) are discrete due to finite $R$. (b) Schematics for possible quantum deviation from classical CVE in scales of $\mu \lesssim \hbar v_F/R$. \label{f9}}
\end{figure}

In continuous limit (infinite size $R$), the gap is zero for metals, while a finite $R$ leads to discrete $k_j=j\frac{2\pi}{R}$, thus discrete energy levels (although they are dense) with a minimum separation $\delta \sim \hbar v_F /R$ (Fig.~\ref{f9}(a)). The $\hbar\omega/\delta=\omega R/v_F\ll1$ ensures a negligible inter-band contribution. In addition, convergent expansion of $\omega$ (Eq.~(\ref{eq28})) requires $\omega$ in the range $\frac{\omega}{v_Fk}<1$. Since $k\sim\frac{2m\pi}{R}$, it yields the condition $\omega R/v_F < 2m\pi$, which is also ensured by Eq.~(\ref{eq62}). 

The second scale is a ratio of two energies $\mu/(\hbar v_F/R)$, where $\hbar v_F/R$ can be viewed as the energy quanta for a particle (at the rim) rotating at $v_F$.
\begin{equation}
\begin{split}
\frac{\mu}{(\hbar v_F/R)}~\begin{cases}
~ \gg 1, & \text{classical}\\
~\lesssim 1, & \text{quantum} 
\end{cases}\label{eq63}
\end{split}
\end{equation}
Classical CVE exhibits $j_{\text{CVE}}\sim\mu^2$ \cite{20,53a}, while its valid condition remains vague. We find deviation should occur in the quantum regime. 

The classical regime corresponds to a situation of numerous $\boldsymbol{k}$ states under the $\mu$ (purple in Fig.~\ref{f9}(a)), allowing a continuous limit $\sum_{\boldsymbol{k}} \to \int_{\boldsymbol{k}}$ that leads to $\mu^2$-dependence; however, in quantum regime (green in Fig.~\ref{f9}(a)), $k_{\perp}=0$ makes the main contribution. 

Note that $k_{\perp}=0$ mode is different from $k_{\perp}\neq0$ modes (Sec. 4.1) in terms of spectrum: $k_{\perp}=0$ mode is symmetric with $k_z$, thus it makes no contribution to $j_{z}$. In terms of evolution, the initial $|s,k_x,k_z,k_z{\rangle}$ of $k_x=k_y=0$ will propagate into $|s,n,k_{\perp},k_z{\rangle}$ of $k_{\perp}=0$: enter the $n=0$ branch for $k_z>0$, and enter the $n=-1$ branch for $k_z<0$ (Fig.~\ref{f4}). 

Thus, the $k_{\perp}=0$ mode vanishes the current when $\mu$ touches the energy $\hbar v_F/R$, instead of a smooth decreasing with $\mu^2$. (Fig.~\ref{f9}(b)); on the other hand, for large $\mu$, the outstanding $k_{\perp}=0$ mode is negligible in weight and $j_{\text{CVE}}$ approaches to $\mu^2$-dependence \cite{20,20a,21,22,23,24}. 

In short, quantum theory indicates that the classical formula is valid for small $\omega$ \textit{and} large $\mu$; precisely, $\frac{\omega}{(v_F/R)}\ll1$ and $\frac{\mu}{(\hbar v_F/R)}\gg 1$. Notably, the valid condition for $j_{\text{CVE}}$ formula establishes links between the system's chemical potential $\mu$, rotation energy quanta $\hbar\omega$, fermi velocity $v_F$ and system's size $R$, which had appeared independent.

\subsection{6. Velocity-independent pumping \& Flat band limit.}
In this section, we discuss the CVE relevance to Fermi velocity $v_F$. We show an implicit indication by Eq.~(\ref{eq32}): whether the Weyl fermion is fast or inert, the amount of charge being pumped by rotation of $2\pi$ is invariant.

In low-$T$ quantum regime, a formula can be obtained based on Eq.~(\ref{eq32})
\begin{equation}
\begin{split}
j_{\text{CVE}}/(-e)=\frac{1}{2}\omega\frac{\mu_+^2-\mu_-^2}{(2\pi)^2v_F^2\hbar^2}=\frac{1}{2}\omega\frac{(k_F^+)^2-(k_F^-)^2}{(2\pi)^2}.\label{eq64}
\end{split}
\end{equation}
By rotating $2\pi$, the pumping charge through an area $A$ is ${\Delta Q}=A{\cdot}j_{\text{CVE}}{\cdot}\frac{2\pi}{\omega}$. Plugging in $k_F^2=(6\pi^2n)^{\frac{2}{3}}$ for 3D isotropic Weyl fermion,
\begin{equation}
\begin{split}
\Delta Q/(-e)=A\left(\frac{9\pi}{16}\right)^{\frac{1}{3}}(n_+^{\frac{2}{3}}-n_-^{\frac{2}{3}})\approx A\left(\frac{9\pi \tilde{n}^2}{16}\right)^{\frac{1}{3}},\label{eq65}
\end{split}
\end{equation}
where $n_{\pm}$ are the volume densities of carriers at each Weyl node. When carriers are mainly concentrated around one node ($n_-\to0$), the $\Delta Q$ is approximately expressed with the \textit{chiral carrier density} $\tilde{n}:=n_+-n_-$. Note that $\tilde{n}$ is defined as the difference between two chiralities $\lambda$, which tends to be zero when Weyl semi-metal is in equilibrium.

The independence of $\Delta Q$ on Fermi velocity reflects that the chiral current driven by rotation is different from the transport driven by electric fields; as in the later case, a flatter band typically decreases the capacity of transport. By using $j=\Delta Q/(A\Delta t)= \tilde{n}{\cdot}v_w=\tilde{n}\frac{\ell_w}{\Delta t}$, we find the shift velocity and path (over $2\pi$-rotation) are $v_F$-independent too. 
\begin{equation}
\begin{split}
v_{w}&=\frac{\omega}{4}(\frac{9}{2\pi^2 \tilde{n}})^{\frac{1}{3}},\\
\ell_w&=(\frac{9\pi}{16 \tilde{n}})^{\frac{1}{3}}.\label{eq66}
\end{split}
\end{equation}

The shift velocity $v_w$ should not be confused with the band velocity $\frac{1}{\hbar}\partial_{\boldsymbol{k}}\epsilon$ which is for an individual particle and proportional to $v_F$. The shift velocity $v_w$ is the net contribution averaged to each chiral carrier -- thus it characterizes the shift of the \textit{holistic} center. For example, consider there are equal amount of particles moving in opposite directions with $v_F$: the band velocity for each particle could be fast $\sim v_F$, while $v_w=0$. Similarly, the shift path $\ell_w:=v_w{\cdot}\frac{2\pi}{\omega}$ is independent of $v_F$. It is also holistic, as it characterizes the shift distance of the charge center over $2\pi$ rotation.
\begin{figure}
\centering
\includegraphics[scale=0.5]{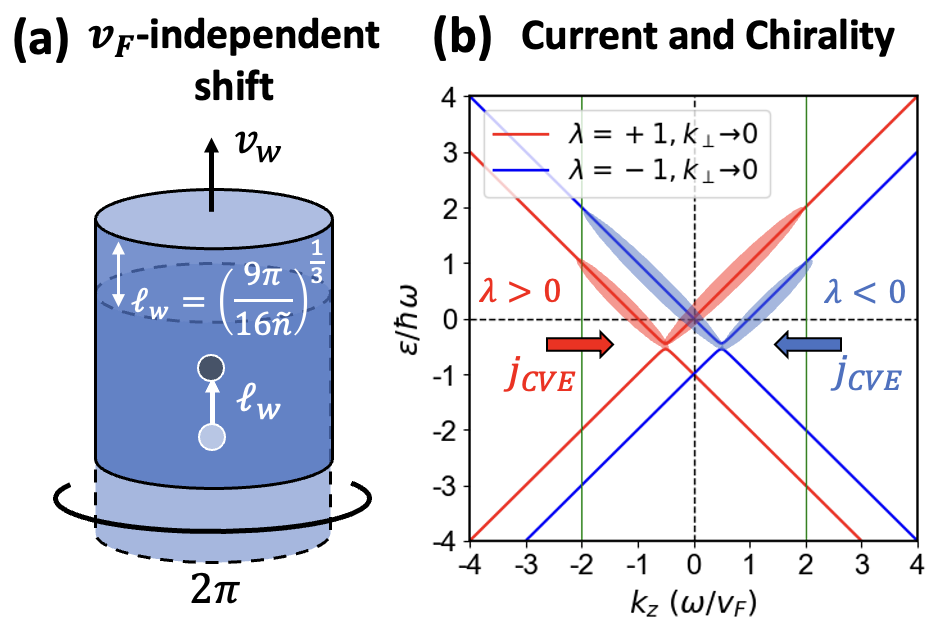}
\caption{\label{fig:epsart}(color online): (a) The shift $\ell_w$ of charge center by rotating $2\pi$. (b) $j_{\text{CVE}}$ flips with chirality $\lambda$ (demonstrated with $\omega>0$, $k_F=2$).   \label{f10}}
\end{figure}

An interesting limit is $v_F\to0$. The shift velocity of the collective charge center (at a given $\omega$) is unaffected even if the individual velocity tends to infinitely slow. On the other hand, if $\mu$ is constant (e.g., via injection of carriers $\tilde{n}$ or continuous chiral anomaly between opposite Weyl nodes), a slower $v_F$ tends to enhance the CVE. This counter-intuitive result can be justified by two points: the velocity-independent pumping and the density of states are increased by a flat band.

However, we should keep in mind two complexities. First, the pre-condition of our treatment: $\omega R \ll v_F$; the analysis is less reliable when $v_F$ decreases to the range $<\omega R$, when inter-band transitions will come into play. Second, the present treatment ignores interaction, while the flat band may invite correlation. \cite{74,75} 

\subsection{7. Role of Berry curvature in chiral transport.}
In this section, we discuss the role of Berry curvature in the CVE, as it is an essential ingredient for understanding the chiral anomaly, CME, etc. \cite{6,24,26,66,67,68}

In semi-classical frameworks, Berry curvature $\boldsymbol{\Omega}$ enters as a classical field of variables $\lbrace\boldsymbol{p},\boldsymbol{x}\rbrace$, due to a velocity $-\dot{\boldsymbol{k}}\times\boldsymbol{\Omega}$ recognized in wave-packet dynamics \cite{27,28,29,30}. The curvature term violates incompressibility suggested by Louville's theorem \cite{18,46,51}; one way to recover that, the volume $\Gamma$ is re-scaled as $d\Gamma=(1+\boldsymbol{B}\cdot\boldsymbol{\Omega})d\boldsymbol{p}d\boldsymbol{x}$ \cite{18,46}. Consequently, the flow $\boldsymbol{j}$ (through a volume) gains the $\boldsymbol{\Omega}$-dependence. Indeed, the CME has been attributed to such curvature-dependent currents \cite{24}
\begin{equation}
\begin{split}
\frac{\boldsymbol{j}(\boldsymbol{x})}{(-e)}=\int_{\boldsymbol{p}}\left(-\epsilon_{\boldsymbol{k}}\frac{\partial n}{\partial \boldsymbol{p}}-\boldsymbol{\Omega}\cdot\frac{\partial n}{\partial\boldsymbol{p}}\epsilon_{\boldsymbol{p}}\boldsymbol{B}-\epsilon_{\boldsymbol{p}}\boldsymbol{\Omega}\times\frac{\partial n}{\partial \boldsymbol{x}}\right).\label{eq67}
\end{split}
\end{equation}

In quantum context, the variable of Berry curvature $\boldsymbol{\Omega}(\boldsymbol{k})$ is merely $\boldsymbol{k}$, and it arises from eigenstates in Hilbert space. The Eq.~(\ref{eq29}) is re-written in terms curvature as (in unit $(-e/\hbar^2)$)
\begin{equation}
\begin{split}
\boldsymbol{j}_{\text{ch}}&=\boldsymbol{\omega}\frac{\hbar^2}{(2\pi)^3}{\sum}_{\lambda,s}{\int}_{\boldsymbol{k}}\lambda s|a_{\lambda,s,\boldsymbol{k}}|^2\frac{1}{2k}\\
&=\boldsymbol{\omega}\frac{\hbar^2}{(2\pi)^3}{\sum}_{\lambda,s}{\int}_{\boldsymbol{k}}|a|^2\lambda s\frac{\hat{\boldsymbol{k}}}{2k^2}{\cdot}\boldsymbol{k}\\
&=\boldsymbol{\omega}\frac{\hbar^2}{(2\pi)^3}{\sum}_{\lambda,s}{\int}_{\boldsymbol{k}}|a|^2(-\boldsymbol{\Omega}_{\lambda,s,\boldsymbol{k}}\cdot\boldsymbol{k}),\label{eq68}
\end{split}
\end{equation}
where the curvature near the Weyl node is $\boldsymbol{\Omega}_{\lambda,s,\boldsymbol{k}}=-\lambda s\frac{\hat{\boldsymbol{k}}}{2k^2}$. Notably, the curvature that appears in current is not from a separate geometric contribution, but is contained in the standard band velocity: $\partial_{\boldsymbol{k}}\epsilon$ gives $\frac{\lambda s}{2k}$ which can be re-written in forms of $\boldsymbol{\Omega}\cdot\boldsymbol{k}$.

The magnetization depends on both the eigenvalues and curvature of eigenstates (Eq.~(\ref{eq23})).
%\begin{equation}
%\begin{split}
%\boldsymbol{M}&=\frac{-e}{(2\pi)^3}\sum_{\lambda,s}{\int}_{\boldsymbol{k}}\frac{|a_{\lambda,s,\boldsymbol{k}}|^2}{\hbar}{\epsilon}_{s,\boldsymbol{k}}\boldsymbol{\Omega}_{\lambda,s,\boldsymbol{k}}
%\end{split}
%\end{equation}
The energy contains two terms
\begin{equation}
\begin{split}
\epsilon_{s,\boldsymbol{k}}=\epsilon_0+\epsilon'=s\hbar v_Fk-\boldsymbol{\omega}{\cdot}(\boldsymbol{x}\times\hbar\boldsymbol{k})\label{eq69}
\end{split}
\end{equation}
The $\epsilon_0$ is isotropic, leading to $\int_{\boldsymbol{k}}\epsilon_0\boldsymbol{\Omega}=\int_{\boldsymbol{k}}\epsilon_0\frac{\hat{\boldsymbol{k}}}{2k^2}=0$. Then, only $\epsilon'$ contributes, and $-\boldsymbol{\omega}{\cdot}(\boldsymbol{x}\times\hbar\boldsymbol{k})$ splits into the following two terms
\begin{equation}
\begin{split}
\nabla_{\boldsymbol{x}}&\times\boldsymbol{M}(\boldsymbol{x})\\
&=\frac{\hbar^2}{(2\pi)^3}\sum_{\lambda,s}{\int}_{\boldsymbol{k}}|a|^2((-\boldsymbol{\Omega}\cdot\boldsymbol{k})\boldsymbol{w}+(\boldsymbol{\Omega}\cdot\boldsymbol{\omega})\boldsymbol{k})\\
&=\frac{2}{3}\boldsymbol{\omega}\frac{\hbar^2}{(2\pi)^3}\sum_{\lambda,s}{\int}_{\boldsymbol{k}}|a|^2(-\boldsymbol{\Omega}_{\lambda,s,\boldsymbol{k}}\cdot\boldsymbol{k}).\label{eq70}
\end{split}
\end{equation}
Integrating the second term above amounts to $-1/3$ of the first term, resulting a factor $2/3$. 

Clearly, both $j_{\text{CVE}}$ and $\nabla_{\boldsymbol{x}}\times\boldsymbol{M}$ depend on $\boldsymbol{\Omega}\cdot\boldsymbol{k}$, and this explains their correlation. When chirality $\lambda$ flips, the curvature $\boldsymbol{\Omega}_{\lambda,s,\boldsymbol{k}}=-\lambda s\frac{\hat{\boldsymbol{k}}}{2k^2}$ flips signs, thus both $j_{\text{CVE}}$ and $\boldsymbol{M}$ flip signs.  
This correlated flip is embedded in the opposite shifts in spectrum Fig.~\ref{f10}(b): $\lambda=\pm1$ leads to $\mp k_z$ shifts thus opposite $j_{\text{CVE}}$.

Notably, Berry curvature enters, for instance, the semi-classical Eq.~(\ref{eq20}) via a modified current formula which includes extra terms in addition to the first term of Eq.~(\ref{eq20}) due to a postulated action. Here, the quantum current remains its standard form, but curvature dependence naturally appears as energy is quantized.

\section{V. Summary and Outlook.}
In the classical framework, the CVE is understood by analog with the CME \cite{6,24,26}, with the Lorentz force simply replaced by the Coriolis force as the mechanism of breaking the chiral symmetry \cite{22,23}. However, non-inertial reference frames and fictitious forces lack counterparts in quantum theory. Consequently, the analogous quantum Hamiltonian of CVE was elusive. Moreover, because the CVE involves rotational motion and may therefore be inherently time-dependent, a tractable semiclassical theory has been favored over a fully quantum description.

This work takes a different approach to the classical-to-quantum transition by abandoning the force-based analogy. That is, rather than seeking a ``quantum force" for the fictitious Coriolis, it quantizes rotation via the operator $\boldsymbol{L}$. This perspective leads to the quantum model developed in Sec. 2. In addition, the $H$ need not explicitly contain time (related to spatial isotropy) and therefore less complicated than one might initially expect. Such a strategy of reduction to time-independence is not unprecedented, such as solving the Rabi oscillation \cite{77} (although the procedure is different). In short, this work demonstrates, first, a viable route to constructing a quantum Hamiltonian for the CVE and, second, that the resulting quantum model is solvable.

\textbf{Three conditions for semiclassical results}. A theory gains credibility if it reproduces established results (Sec. 3). More importantly, however, it should offer an extended picture, such as what conditions for semi-classical formula to hold. This work reveals that the semi-classical results rely on three conditions: 

(1) Intra-band approximation: $\omega R/v_F \ll1$.

(2) High-degeneracy approximation: $\mu/(\hbar v_F R)\gg1$.

(3) Isotropic approximation (in the $x$-$y$ plane).

Condition (1) may alternatively be described as the requirement of a ``slow rotation" (small $\omega$), while condition (2) corresponds to a ``large chemical potential" (large $\mu$), both relative to the scales set by $R$ and $v_F$. The three conditions are independent and must be satisfied simultaneously.

How should these conditions be interpreted? For condition (1), a large $\omega$ induces substantial inter-band transition, causing both Weyl and non-Weyl bands to contribute and thereby obscuring a pure CVE response. By contrast, the semiclassical theory assumes no upper bound for $\omega$, effectively corresponding to an infinite band gap. The quantum theory suggests that in Weyl semi-metals, $\omega$ must be capped, depending on the system size and fermi velocity. 

For condition (2), $\mu$ sets the degeneracy $N_{k_{\perp}}$ and thereby determines the relative contribution of $k_{\perp}=0$ modes which are affected by the quantum void states absent from the semiclassical framework. In simple terms, a large $\mu$ dilutes the influence of void states, such that the observable will tend to the semiclassical limit (Fig.~\ref{f9}(b)). 

The significance of condition (3) has often been overlooked because it was initially introduced as a technical assumption to simplify the analysis. However, the effect of anisotropy proves more than using an orientation-resolved distribution $f$ (Sec. 4.3). 

\textbf{CVE: A class of non-equilibrium states}. As shown in Sec. 4.4, the CVE lacks a GS, which distinguishes it fundamentally from the CME. This difference can be traced to the Floquet character of the CVE spectrum: the Weyl band splits into infinite Floquet folds (Fig.~\ref{f2}(b)), a feature absent in the CME. These results suggest that, although the classical analogy between the CVE and CME is highly useful at the semiclassical level, it does not fully capture the distinctive quantum structure of the CVE.

Consequently, the semi-classical formulas Eqs.~(\ref{eq31}) and (\ref{eq43}) should describe a class of metastable states originating from (isotropic) initial states $\hat{f}_0$. The correlations of $\omega$ with observables such as $j_{z}$ or $\boldsymbol{M}(\boldsymbol{x})$ hold whether the state $\hat{f}_0$ is close to or far from equilibrium. The specific $\hat{f}_0$ will influence the response coefficients. From this perspective, the CVE is more appropriately viewed as a family of non-equilibrium states, rather than a unique one generated from the unperturbed GS.

Notably, the term non-equilibrium has two distinct meanings in the present context. The first refers to the population imbalance between two Weyl nodes of opposite chirality in a Weyl semimetal. This imbalance is not itself the source of the novel physics; rather, it serves as a means of realizing a chiral fermion system. The second refers to a distribution within a single Weyl node that deviates from its ground-state (or equilibrium) distribution. The identification of the CVE as a non-equilibrium phenomenon pertains to this second meaning. 

\textbf{Is there an annealing process or relaxation time scale associated with CVE ?} From the onset of rotation to the establishment of $j_{\text{CVE}}$, does the system undergo a quantum annealing or dephasing process? In other words, is the distribution $f_{\text{CVE}}$ is formed by a unitary dynamical evolution, or does it involve a thermodynamical relaxation characterized by a finite time scale \cite{1,2,26}? Existing semiclassical approaches appear incapable in addressing this, since the property of $f_{\text{CVE}}$ is assumed from the beginning.

The present results offer important clues. The evolution generated by $H$ (Sec. 4.2) is a standard unitary $U(t)$ and involves neither annealing nor dephasing. Based on that, the established CVE phenomena (e.g., currents, magnetization) are readily reproduced. This suggests that the distribution $f_{\text{CVE}}(\epsilon)$ emerges from the dynamical evolution following the introduction of $\hat{V}$. Thus, the CVE is more naturally understood as a non-equilibrium dynamical effect than a thermodynamical one. 

Conversely, if relaxation is an intrinsic ingredient of the CVE, existing theories \cite{1,2,3,4,5,8,18,19,20,20a,21,22,23,24} may need to be extended to incorporate a characteristic relaxation timescale. Such a scenario would also raise a conceptual challenge: whether a thermal distribution of CVE remains invariant under transformations between different reference frames of either Lorentz or Galilean types.

\textbf{Conceptual progress facilitates experimental observation}. The quantum model indicates that rotating the orbital degrees of freedom $\hat{V}=-\omega L_z$ leads to $j_{\text{CVE}}$; on the other hand, rotating both spin and orbital degrees freedom (i.e., $\hat{V}=-\omega J_z$) only leads to $j_z=0$. Because $[-\omega J_z,H_0]=0$, the observable ${\langle}j_z{\rangle}=0$ is conserved under $\hat{V}=-\omega J_z$. In other words, operator $J_z$ generates the reference framework transformation, which affects ${\omega}_0$, instead of $\omega$. Physically, that means, the essential meaning of rotation in CVE is the relative motion between the orbital and spin degrees of freedom. 

Unlike CME which can readily be controlled by magnetic field \cite{26,37}, the CVE is difficult to realize steady matter rotation on lattice (because charge is strongly bound to the lattice). The present theory suggests a simpler equivalent scheme based on rotating spin. We do not have to separate charge from the lattice but just need to generate a spin rotation of $-\omega$. This will generate an equivalent CVE observable just as charge is rotating with $\omega$. 

\textbf{About the nature of observables and the description of rotation}. Regarding the phenomena examined in this work, we emphasize that they are bulk in nature because the Hamiltonian considered here describes an infinite system without boundaries. While the current density $j_z$ is numerically equal to the on-axis current density \cite{1}, it represents a current distributed throughout the volume and is therefore a bulk observable \cite{22,24,46,53a}, conceptually distinct from a current localized on the rotation axis \cite{17} or from effects arising from boundary conditions \cite{8a}.

In describing rotation in the CVE \cite{17,8a,53a}, the total angular momentum $J_z$ or the orbital angular momentum $L_z$ may serve as the generator. Since $[J_z, H_0]=0$, the influence by $\hat{V}=\omega J_z$ produces no bulk current density because $H$ and $H_0$ share the common eigenstates. A non-zero current can arise only precisely on the axis $r=0$ \cite{1,2} or in a finite system, as a consequence of boundary conditions \cite{8a}. In contrast, $[L_z, H_0]\neq0$, and $\hat{V}=\omega L_z$ induces a finite bulk current density and magnetization (Sec. 3). This work focuses on the case of $\hat{V}=\omega L_z$ and finds a non-zero bulk-averaged current density which is experimentally accessible.

\textbf{Conclusion}.
The CVE is presently defined as a charge current response to rotation, understood within semiclassical frameworks based on distribution $f$. A central achievement of this work is developing a quantum formulation and identifying the microscopic complete wavefunction underlying $f$. The state resolves both the phase and spin degrees of freedom, coherently connecting semiclassical results in different aspects (Secs. 3.1, 3.2). Beyond semiclassical scopes, it uncovers the void states, a distinct space structure from classical states (Sec. 4.1); reveal the roles of isotropic approximation and the inadequacy of semiclassical methods in anisotropic situations (Secs. 4.2, 4.3); evaluates the distribution $f(\epsilon)$ in different reference frames and quantitatively tests the semiclassical assumption (Sec. 4.4); uncovers two quantum scales and the crossover conditions (Sec. 4.5); Predicts an axial charge pumping due to a holistic shift velocity, enhanced by a decreased $v_F$ (Sec. 4.6); illustrates the role Berry curvature $\boldsymbol{\Omega}$ in chiral transport (Sec. 4.7).

\textbf{Outlook}. Thus far, we have employed the spinful wavefunction only to evaluate orbital observables (e.g, $j_z$, orbital magnetization $\boldsymbol{M}$), both within and beyond the semiclassical regime. % This is because...

An intriguing next question is whether rotation drives spin dynamics in parallel with axial charge transport and, if so, whether these responses are correlated. With the spin-resolved wavefunction now established, the scope of the CVE may extend beyond charge transport to include spin-dependent observables. 

%The quantum theory presented here renders such questions quantitatively testable. In this sense, obtaining the full quantum Hamiltonian and wavefunction is not merely a formal achievement but also a prerequisite for exploring new classes of rotational responses. 

More broadly, the spin-resolved wavefunction provides an opportunity to reassess assumptions that have traditionally been made in CVE studies. In particular, to what extent can a spinless distribution $f(\boldsymbol{x},\boldsymbol{p})$ handle the problem without loss of essential physics?

\textbf{Acknowledgement}. We wish to acknowledge the inspiring discussion with Pavan Hosur, Bishnu Karki, Jonathan D. H. Smith, Jigang Wang in the course of preparing this work. This work was supported by the Department of Energy grant number DE-SC0022264.

\appendix

\section{Appendix}
\subsection{A. From plane waves to Bessel functions.}
Here, we derive inner product $\langle s,n,k_{\perp},k_z|s,k_x,k_y,k_z \rangle$ between the plane-wave and Bessel function bases.
The transition coefficients are defined as
\begin{equation}
\begin{split}
|s,n,k_{\perp},k_z{\rangle}&={\sum}d_{s',m,k_{\perp}',k_z'}^{(s,n,k_{\perp},k_z)}|s',\delta_m,k_{\perp}',k_z'{\rangle},\\
|s',\delta_m,k_{\perp}',k_z'{\rangle}&={\sum}c_{s,n,k_{\perp},k_z}^{(s',m,k_{\perp}',k_z')}|s,n,k_{\perp},k_z{\rangle},\label{eq71}
\end{split}
\end{equation}
Given a small $\omega$ (the intra-band approximation in Sec. 5), it reduces to
\begin{equation}
\begin{split}
d_{s',m,k_{\perp}',k_z'}^{(s,n,k_{\perp},k_z)}&=\delta_{s,s'}\delta_{k_z,k_z'}\delta_{k_{\perp},k_{\perp}'}{\cdot}d_{m,k_{\perp}}^{(n)},\\
c_{s,n,k_{\perp},k_z}^{(s',m,k_{\perp}',k_z')}&=\delta_{s,s'}\delta_{k_z,k_z'}\delta_{k_{\perp},k_{\perp}'}{\cdot}c_{n,k_{\perp}}^{(m)}.\label{eq72}
\end{split}
\end{equation}
Here, $s$ labels the band, and the factor $\delta_{s,s'}$ indicates that inter-band transitions $s\to s'$ are negligible after the perturbation is applied. Since $|k_z\rangle$ forms an independent quotient space, reduced coefficients could be defined as
\begin{equation}
\begin{split}
d_{m,k_{\perp}}^{(n)}&={\langle}s,\delta_m,k_{\perp},k_z|s,n,k_{\perp},k_z{\rangle}={\langle}s,\delta_m,k_{\perp}|s,n,k_{\perp}{\rangle},\\
c_{n,k_{\perp}}^{(m)}&={\langle}s,n,k_{\perp},k_z|s,\delta_m,k_{\perp},k_z{\rangle}={\langle}s,n,k_{\perp}|s,\delta_m,k_{\perp}{\rangle}.\label{eq73}
\end{split}
\end{equation}
The coefficients $a_{\pm}^{(s)}(\boldsymbol{k})$ and $b_{\pm}^{(s)}(k_{\perp},k_z)$ in the eigenstates for $H_0$ and $H$:
\begin{widetext}
\begin{equation}
\begin{split}
\langle s_z,r_x,r_y,r_z |s,\boldsymbol{k}\rangle &=\begin{pmatrix} a_+^{(s)}(k_x,k_y,k_z) \\ a_-^{(s)}(k_x,k_y,k_z) \end{pmatrix} e^{ik_xr_x}e^{ik_yr_y}e^{ik_zr_z} \\
\langle s_z,r,\phi,r_z|s,n,k_{\perp},k_z \rangle &= \begin{pmatrix} b_+^{(s)}(k_{\perp},k_z)J_n(k_{\perp}r)e^{in\phi} \\ b_-^{(s)}(k_{\perp},k_z)J_{n+1}(k_{\perp}r)e^{i(n+1)\phi} \end{pmatrix}e^{ik_zr_z}\label{eq74}
\end{split}
\end{equation}
Formally, the coefficients are determined by the integral
\begin{equation}
\begin{split}
c_{n,k_{\perp}}=\sum_l{\int}a_{l}^{(s)*}e^{-i(k_xr_x+k_yr_y)}b_{l}^{(s)}J_{n+{\Theta_l}}(k_{\perp}r)e^{i(n+\Theta_l)\phi}d{\phi}{\cdot}rdr,\label{eq75}
\end{split}
\end{equation}
where $l=\pm$, $\Theta_{+}=0$, $\Theta_{-}=1$, and the variables $\boldsymbol{k}$ or $k_{\perp}$ of $a^{(s)},b^{(s)}$ are neglected. 
\end{widetext}

However, in the continuous limit, Eq.~(\ref{eq75}) must be divergent. To demonstrate this, we assume Eq.~(\ref{eq75}) is convergent and then examine the inconsistency caused by this assumption. For convenience, $|s,k_x,k_y{\rangle}:=|s,\delta,k_{\perp}{\rangle}$, with $k_{\perp}\text{cos}{\delta}=k_x$ and $k_{\perp}\text{sin}{\delta}=k_y$. We also define
\begin{equation}
\begin{split}
[\varphi_{\delta,k_{\perp}}^{(s)}(\boldsymbol{r})]_l&:=a_{l}^{(s)}e^{-i(k_xr_x+k_yr_y)}=a_{l}^{(s)}e^{-ik_{\perp}(r_x\text{cos}{\delta}+r_y\text{sin}\delta)}\\
&[\psi_{n,k_{\perp}}^{(s)}(r,\phi)]_l:=b_{l}^{(s)}J_{n+{\Theta_l}}(k_{\perp}r).\label{eq76}
\end{split}
\end{equation}
These functions satisfy
\begin{equation}
\begin{split}
\varphi_{\delta,k_{\perp}}^{(s)}(\boldsymbol{r})=\varphi_{\delta,1}^{(s)}(k_{\perp}\boldsymbol{r}),~\psi_{n,k_{\perp}}^{(s)}(r,\phi)=\psi_{n,1}^{(s)}(k_{\perp}r,\phi)\label{eq77}
\end{split}
\end{equation}
We examine the scaling behavior of $c_{n,k_{\perp}}$ with $k_{\perp}$. Start from unity $k_{\perp}=1$,
\begin{equation}
\begin{split}
c_{n,1}=\sum_l{\int}_{\infty}[\varphi_{\delta,1}^{*}(\boldsymbol{r})]_l[\psi_{n,1}(r)]_l{\cdot}e^{i(n+\Theta_j)\phi}d{\phi}{\cdot}rdr,\label{eq78}
\end{split}
\end{equation}
Rescale the variable $\boldsymbol{r}':=k_{\perp}\boldsymbol{r}$ with $k_{\perp}>0$ and $k_{\perp}\neq1$. Since the integration is over an infinite rangle, scaling $k_{\perp}$ does not change the form of integration 
\begin{equation}
\begin{split}
&c_{n,k_{\perp}}=\sum_l{\int}_{\infty}[\varphi_{\delta,k_{\perp}}^{*}(\boldsymbol{r})]_l[\psi_{n,k_{\perp}}(r)]_l{\cdot}e^{i(n+\Theta_j)\phi}d{\phi}{\cdot}rdr\\
&=\sum_l{\int}_{\infty}[\varphi_{\delta,1}^{*}(k_{\perp}\boldsymbol{r})]_l[\psi_{n,1}(k_{\perp}r)]_l{\cdot}e^{i(n+\Theta_j)\phi}d{\phi}{\cdot}rdr\\
&=\frac{1}{k_{\perp}^2}\sum_j{\int}_{\infty}[\varphi_{\delta,1}^{*}(\boldsymbol{r}')]_l[\psi_{n,1}(r')]_le^{i(n+\Theta_j)\phi}d{\phi}{\cdot}r'dr'\\
&=\frac{1}{k_{\perp}^2}c_{n,1}.\label{eq79}
\end{split}
\end{equation}
Equation~(\ref{eq79}) gives the relationship between $c_{n,k_{k_{\perp}}}$ and $c_{n,1}$. In addition, the relationship can be deduced from the fact that the density of mesh $(k_x,k_y)$ is the same under cartesian coordinates, we have (the magnitude depends on the length of ring $2{\pi}k_{\perp}$)
\begin{equation}
\begin{split}
|c_{n,k_{\perp}}/c_{n,1}|^2=1/k_{\perp}.\label{eq80}
\end{split}
\end{equation}
Combine Eq.~(\ref{eq79}) and Eq.~(\ref{eq80}), we find (for $k_{\perp}>0$ and $k_{\perp}\neq1$)
\begin{equation}
\begin{split}
\frac{|c_{n,k_{\perp}}|}{\sqrt{k_{\perp}}}(1-k_{\perp}^{-\frac{3}{2}})=0.\label{eq81}
\end{split}
\end{equation}
The only solution is $c_{n,k_{\perp}}{\equiv}0$. That means if Eq.~(\ref{eq75}) is convergent, the wavefunction is constantly vanishing, evidently meaningless. Thus, the assumed integration cannot be convergent with $\int_{\infty}d\boldsymbol{r}$. That means, alternative definition is needed for the inner product. 

Next, we find the convergent definition and illustrate why divergence has happened. The structure of eigenbases is $|s,n,k_{\perp},k_z{\rangle}=|s,n,k_{\perp}{\rangle}{\otimes}|k_z{\rangle}$. Since $|k_z{\rangle}$ depends on $r_z$, we should only concern about $|s,n,k_{\perp}{\rangle}$
\begin{equation}
\begin{split}
|s,n,k_{\perp}{\rangle}=\sum_{m}^{N_{k_{\perp}}}d_{m,k_{\perp}}^{(n)}|s,\delta_m,k_{\perp}{\rangle},\label{eq82}
\end{split}
\end{equation}
where $\delta_m$ is the orientation angle of a $k_x$-$k_y$ vector. The sum cover $N_{k_{\perp}}$ vectors $(k_x,k_y)$ located in the ring of radius $k_{\perp}$, each of which corresponds a polar angle $\delta_m$. Both the left and right sides are eigenstates of operator $\hat{k}_x^2+\hat{k}_y^2$. Thus, inner products satisfy ${\langle}k_{\perp}|k_{\perp}'{\rangle}=\delta_{k_{\perp},k_{\perp}'}$. 

Apply a rotation operation $\hat{R}(\phi_0)$ to the coordinate frame,
\begin{equation}
\begin{split}
\hat{R}(\phi_0)|s,n,k_{\perp}{\rangle}&=e^{i(n+\frac{1}{2})\phi_0}|s,n,k_{\perp}{\rangle}\\
&=\sum_{m}d_{m,k_{\perp}}^{(n)}\hat{R}(\phi_0)|s,\delta_m,k_{\perp}{\rangle}.\label{eq83}
\end{split}
\end{equation}
With $\phi_0=2\pi\frac{m_0}{N_{k_{\perp}}}$, we have $\hat{R}(\phi_0)|\delta_m,k_{\perp}{\rangle}{\sim}|\delta_m+\phi_0,k_{\perp}{\rangle}=|\delta_{m+m_0},k_{\perp}{\rangle}$, and
\begin{equation}
\begin{split}
|s,n,k_{\perp}{\rangle}=\sum_{m}e^{-i(n+\frac{1}{2})\phi_0}d_{m,k_{\perp}}^{(n)}|s,\delta_{m+m_0},k_{\perp}{\rangle}.\label{eq84}
\end{split}
\end{equation}
Compared with Eq.~(\ref{eq82}), we obtain for arbitrary $m_0$
\begin{equation}
\begin{split}
d_{m,k_{\perp}}^{(n)}&=e^{-i(n+\frac{1}{2})\phi_0}d_{m+m_0,k_{\perp}}^{(n)}\\
&~{\Rightarrow}~|d_{m,k_{\perp}}^{(n)}|=|d_{m+m_0,k_{\perp}}^{(n)}|=1/\sqrt{N_{k_{\perp}}}.\label{eq85}
\end{split}
\end{equation}
Selecting a gauge $d_{0,k_{\perp}}^{(n)}=e^{i\pi\frac{m}{N_{k_{\perp}}}}$ (i.e., adsorbing this $m$-dependent phase to bases $|s,\delta_m,k_{\perp}\rangle$), we can express the coefficients as
\begin{equation}
\begin{split}
d_{m,k_{\perp}}^{(n)}=\frac{1}{\sqrt{N_{k_{\perp}}}}e^{i2\pi(nm)/{N_{k_{\perp}}}}.\label{eq86}
\end{split}
\end{equation}
An inverse transformation (a Fourier type) leads to 
\begin{equation}
\begin{split}
c_{n,k_{\perp}}^{(m)}&=\frac{1}{\sqrt{N_{k_{\perp}}}}e^{-i2\pi(nm)/{N_{k_{\perp}}}}\\
|s,\delta_m,k_{\perp}{\rangle}&=\sum_{n}^{N_{k_{\perp}}}\frac{1}{\sqrt{N_{k_{\perp}}}}e^{-i2\pi(nm)/{N_{k_{\perp}}}}|s,n,k_{\perp}{\rangle}\label{eq87}
\end{split}
\end{equation}
Since the expectation value of $J_z$ for the plane wave is zero (the initial state is not rotating), thus the integer $n$ should be symmetric in positive and negative values.

Clearly, Eq.~(\ref{eq87}) justifies why the scaling analysis in the continuous limit directly leads to vanishing wavefunction. Because the continuous limit directly makes $N_{k_{\perp}}\to\infty$ and thus $c_{n,k_{\perp}}{\to}0$.

\subsection{B. Summary of notations.}
We summarize the denotations adopted by this work.

$\lambda$: Chirality.

$\mu$: Chemical potential.

$v_F$: Fermi velocity. 

$\boldsymbol{J}$, $J_z$: Total angular momentum.

$\boldsymbol{L}$, $L_z$: Orbital angular momentum.

$\boldsymbol{j}$, $j_z$: Charge current density (\textit{not} particle current).

$\omega_0$: Angular speeds of the observer.

$\omega_{\text{f}}$: Angular speeds of the Weyl fermion. 

$\omega_{\text{l}}$: Angular speeds of the lattice.

$N_{k_{\perp}}$: Degeneracy of states at $k_{\perp}=\sqrt{k_x^2+k_y^2}$.

$R$: The radius of the system (in the $x$-$y$ plane).

$\boldsymbol{M}$: Magnetization (orbital). 

$\boldsymbol{M}_{\text{IC}}$, $\boldsymbol{M}_{\text{LC}}$: Itinerant and localized parts of orbital magnetization. 

$\boldsymbol{\Omega}$: Berry curvature.

$\boldsymbol{r}$ ($r_x,r_y,r_z$ or $r,\phi,r_z$): The position coordinates.

$\boldsymbol{x}$, $\nabla_{\boldsymbol{x}}$: The coarse spatial coordinates and derivative.

$\boldsymbol{k}=\boldsymbol{p}/\hbar$ ($k_x,k_y,k_z$ or $k_{\perp}$, $\delta_m$, $k_z$): The momentum.

$e$: Elementary charge (positive-valued).

$\hat{f}$: The density operator.

$f_F(\epsilon)$: Fermi distribution function.

$\mathcal{C}[...]$, $\mathcal{Q}[...]$: Classical and quantum reference frame transformations operations.

$\langle s, k_x,k_y,k_z|$: Eigenstates of $H_0$, where $s$ is the band label.

$\langle s, n,k_{\perp},k_z|$: Eigenstates of $H$, where $n$ is an integer quantum number of $J_z$, and $k_{\perp}$ is for operator $k_x^2+k_y^2$.

$a_{\pm}^{(s)}(k_x,k_y,k_z)$ and $b_{\pm}^{(s)}(k_{\perp},k_z)$: Coefficients in the eigenfunctions of $H_0$ or $H$ (see in Secs. 2.1, 4.1). 

$a_{s,k_x,k_y,k_z}$ and $b_{s,n,k_{\perp},k_z}$: Coefficients of the density operators (see in Secs. 4.2, 4.3).

$d_{s',m,k_{\perp}',k_z'}^{(s,n,k_{\perp},k_z)}$ and $c_{s,n,k_{\perp},k_z}^{(s',m,k_{\perp}',k_z')}$: Coefficients of basis transformation (see in Appendix A).

\end{document}